\documentclass[twocolumn]{aastex631}

\hypersetup{linkcolor=magenta,citecolor=blue,filecolor=cyan,urlcolor=purple}

\usepackage{graphicx}
\usepackage[caption=false]{subfig}
\usepackage{amsmath}

\usepackage{array,booktabs,threeparttable}
\usepackage{tabularx}

\received{October 25, 2023}
\revised{December 20, 2023}
\accepted{January 2, 2024}
\submitjournal{ApJ}

\shorttitle{Mesoscale Structure of CMEs at Mercury's Orbit}
\shortauthors{Palmerio et al.}

\graphicspath{{./}{figures/}}

\turnoffeditone
\turnoffedittwo
\turnoffeditthree

\begin{document}

\title{On the Mesoscale Structure of CMEs at Mercury's Orbit:\\ BepiColombo and Parker Solar Probe Observations}

\correspondingauthor{Erika Palmerio}
\email{epalmerio@predsci.com}

\author[0000-0001-6590-3479]{Erika Palmerio}
\affiliation{Predictive Science Inc., San Diego, CA 92121, USA}

\author[0000-0003-1758-6194]{Fernando Carcaboso}
\affiliation{Postdoctoral Program Fellow, NASA Goddard Space Flight Center, Greenbelt, MD 20771, USA}

\author[0000-0003-0412-1064]{Leng Ying Khoo}
\affiliation{Department of Astrophysical Sciences, Princeton University, Princeton, NJ 08544, USA}

\author[0000-0001-6813-5671]{Tarik M. Salman}
\affiliation{Heliophysics Science Division, NASA Goddard Space Flight Center, Greenbelt, MD 20771, USA}
\affiliation{Department of Physics and Astronomy, George Mason University, Fairfax, VA 22030, USA}

\author[0000-0003-0277-3253]{Beatriz S{\'a}nchez-Cano}
\affiliation{School of Physics and Astronomy, University of Leicester, LE1 7RH Leicester, UK}

\author[0000-0001-6886-855X]{Benjamin J. Lynch}
\affiliation{Space Sciences Laboratory, University of California, Berkeley, CA 94720, USA}
\affiliation{Department of Earth, Planetary, and Space Sciences, University of California, Los Angeles, CA 90095, USA}

\author[0000-0002-8748-2123]{Yeimy J. Rivera}
\affiliation{Center for Astrophysics, Harvard \& Smithsonian, Cambridge, MA 02138, USA}

\author[0000-0002-6302-438X]{Sanchita Pal}
\affiliation{Postdoctoral Program Fellow, NASA Goddard Space Flight Center, Greenbelt, MD 20771, USA}

\author[0000-0003-0565-4890]{Teresa Nieves-Chinchilla}
\affiliation{Heliophysics Science Division, NASA Goddard Space Flight Center, Greenbelt, MD 20771, USA}

\author[0000-0002-6273-4320]{Andreas J. Weiss}
\affiliation{Postdoctoral Program Fellow, NASA Goddard Space Flight Center, Greenbelt, MD 20771, USA}

\author[0000-0002-3176-8704]{David Lario}
\affiliation{Heliophysics Science Division, NASA Goddard Space Flight Center, Greenbelt, MD 20771, USA}

\author[0000-0002-7539-0803]{Johannes Z. D. Mieth}
\affiliation{Institut f{\"u}r Geophysik und extraterrestrische Physik, TU Braunschweig, D-38106 Braunschweig, Germany}

\author[0000-0001-7894-8246]{Daniel Heyner}
\affiliation{Institut f{\"u}r Geophysik und extraterrestrische Physik, TU Braunschweig, D-38106 Braunschweig, Germany}

\author[0000-0002-7728-0085]{Michael L. Stevens}
\affiliation{Center for Astrophysics, Harvard \& Smithsonian, Cambridge, MA 02138, USA}

\author[0000-0002-4559-2199]{Orlando M. Romeo}
\affiliation{Space Sciences Laboratory, University of California, Berkeley, CA 94720, USA}
\affiliation{Department of Earth and Planetary Science, University of California, Berkeley, CA 94720, USA}

\author[0000-0002-2542-9810]{Andrei N. Zhukov}
\affiliation{Solar--Terrestrial Centre of Excellence---SIDC, Royal Observatory of Belgium, B-1180 Brussels, Belgium}
\affiliation{Skobeltsyn Institute of Nuclear Physics, Moscow State University, 119991 Moscow, Russia}

\author[0000-0002-6097-374X]{Luciano Rodriguez}
\affiliation{Solar--Terrestrial Centre of Excellence---SIDC, Royal Observatory of Belgium, B-1180 Brussels, Belgium}

\author[0000-0002-1604-3326]{Christina O. Lee}
\affiliation{Space Sciences Laboratory, University of California, Berkeley, CA 94720, USA}


\author[0000-0002-0978-8127]{Christina M. S. Cohen}
\affiliation{California Institute of Technology, Pasadena, CA 91125, USA}

\author[0000-0003-2361-5510]{Laura Rodr{\'i}guez-Garc{\'i}a}
\affiliation{European Space Astronomy Centre, European Space Agency, E-28692 Villanueva de la Ca\~{n}ada, Madrid, Spain}
\affiliation{Universidad de Alcal{\'a}, Space Research Group (SRG-UAH), E-28801 Alcal{\'a} de Henares, Madrid, Spain}

\author[0000-0002-7287-5098]{Phyllis L. Whittlesey}
\affiliation{Space Sciences Laboratory, University of California, Berkeley, CA 94720, USA}

\author[0000-0003-3903-4649]{Nina Dresing}
\affiliation{Department of Physics and Astronomy, University of Turku, FI-20014 Turku, Finland}

\author[0000-0003-0794-7742]{Philipp Oleynik}
\affiliation{Department of Physics and Astronomy, University of Turku, FI-20014 Turku, Finland}

\author[0000-0002-0606-7172]{Immanuel C. Jebaraj}
\affiliation{Department of Physics and Astronomy, University of Turku, FI-20014 Turku, Finland}

\author[0000-0002-8435-7220]{David Fischer}
\affiliation{Space Research Institute, Austrian Academy of Sciences, A-8042 Graz, Austria}

\author[0000-0001-7818-4338]{Daniel Schmid}
\affiliation{Space Research Institute, Austrian Academy of Sciences, A-8042 Graz, Austria}

\author[0000-0002-5324-4039]{Ingo Richter}
\affiliation{Institut f{\"u}r Geophysik und extraterrestrische Physik, TU Braunschweig, D-38106 Braunschweig, Germany}

\author[0000-0003-1411-217X]{Hans-Ulrich Auster}
\affiliation{Institut f{\"u}r Geophysik und extraterrestrische Physik, TU Braunschweig, D-38106 Braunschweig, Germany}

\author[0000-0002-5456-4771]{Federico Fraschetti}
\affiliation{Center for Astrophysics, Harvard \& Smithsonian, Cambridge, MA 02138, USA}
\affiliation{Department of Planetary Sciences, Lunar and Planetary Lab, Tucson, AZ 85721, USA}

\author[0000-0003-4105-7364]{Marilena Mierla}
\affiliation{Solar--Terrestrial Centre of Excellence---SIDC, Royal Observatory of Belgium, B-1180 Brussels, Belgium}
\affiliation{Institute of Geodynamics of the Romanian Academy, 020032 Bucharest-37, Romania}


\begin{abstract}

On 2022 February 15, an impressive filament eruption was observed off the solar eastern limb from three remote-sensing viewpoints, namely Earth, STEREO-A, and Solar Orbiter. In addition to representing the most-distant observed filament at extreme ultraviolet wavelengths---captured by Solar Orbiter’s field of view extending to above 6\,$R_{\odot}$---this event was also associated with the release of a fast ($\sim$2200~km$\cdot$s$^{-1}$) coronal mass ejection (CME) that was directed towards BepiColombo and Parker Solar Probe. These two probes were separated by 2$^{\circ}$ in latitude, 4$^{\circ}$ in longitude, and 0.03~au in radial distance around the time of the CME-driven shock arrival in situ. The relative proximity of the two probes to each other and to the Sun ($\sim$0.35~au) allows us to study the mesoscale structure of CMEs at Mercury’s orbit for the first time. We analyse similarities and differences in the main CME-related structures measured at the two locations, namely the interplanetary shock, the sheath region, and the magnetic ejecta. We find that, despite the separation between the two spacecraft being well within the typical uncertainties associated with determination of CME geometric parameters from remote-sensing observations, the two sets of in-situ measurements display some profound differences that make understanding of the overall 3D CME structure particularly challenging. Finally, we discuss our findings within the context of space weather at Mercury’s distances and in terms of the need to investigate solar transients via spacecraft constellations with small separations, which has been gaining significant attention during recent years.

\end{abstract}



\section{Introduction} \label{sec:intro}

Coronal mass ejections (CMEs) can be enumerated amongst the most spectacular and energetic events in the solar system, consisting of enormous clouds of plasma and magnetic fields that are frequently launched from the Sun into the heliosphere. As they travel (and expand) through the solar corona and interplanetary space, CMEs can experience a myriad of processes, including deflection \citep[e.g.,][]{wang2004, liewer2015, cecere2020}, rotation \citep[e.g.,][]{yurchyshyn2007, vourlidas2011, thompson2012}, erosion \citep[e.g.,][]{dasso2007, ruffenach2012, pal2021}, and deformation \citep[e.g.,][]{savani2010, barnard2017, braga2022} due to e.g.\ interaction with the structured solar wind \citep[e.g.,][]{isavnin2013, heinemann2019, davies2021, maunder2022} or other CMEs \citep[e.g.,][]{lugaz2012, lugaz2014, temmer2014, kilpua2019a}. By the time they reach 1~au, CMEs measure on average ${\sim}$0.3~au in radial size \citep[e.g.,][]{jian2018} and may have lost their coherence as magnetohydrodynamic structures \citep[e.g.,][]{owens2017}. The plethora of phenomena that can act on a CME as it journeys away from the Sun may result in structures encountered in situ that are especially difficult to interpret, and even properties such as geoeffectiveness can be altered significantly during propagation due to these evolutionary processes \citep[e.g.,][]{lavraud2014, mostl2015, liu2018, scolini2020, winslow2021, palmerio2022b}.

Indeed, determination of the global 3D structure of CMEs from in-situ measurements, each representing a single 1D trajectory through a humongous plasma cloud, is not a trivial task even in the case of ``simpler'' situations in which a CME is propagating more or less radially and self-similarly. In this context, multi-point measurements can provide extra constraints on the local structure of a CME at different points throughout its angular extent. The power of multi-spacecraft observations of CMEs in situ has been known for several decades \citep[e.g.,][]{burlaga1981}, and because of data availability over the recent years, multi-probe studies have been gaining increasing attention. However, most multi-point measurements are attained over arbitrary radial and angular separations of the spacecraft involved \citep[e.g.,][]{davies2022, mostl2022, rodriguezgarcia2022}, making it particularly difficult to attribute structural and compositional differences of the investigated CMEs to radial evolution, to longitudinal (local) variations, or to both. Even more so, difficulties only increase as CMEs are probed beyond 1~au and towards the outer heliosphere, mainly due to heightened chances of interactions and mergers of initially-separated structures \citep[e.g.,][]{rodriguez2008, witasse2017, palmerio2021}. Nevertheless, there are some ``special'' cases of spacecraft configurations that are of particular interest for CME studies, i.e.\ those that involve small longitudinal separations, allowing focussed analyses on radial evolution \citep[e.g.,][]{good2019, vrsnak2019, salman2020a}, or small radial separations, permitting investigations of the overall structure at one snapshot in time \citep[e.g.,][]{kilpua2009, farrugia2011, lugaz2022}.

An additional ``special'' case in terms of relative spacecraft configurations takes place when the probes involved are characterised by small longitudinal and radial separations at the same time, thus providing the opportunity to examine the mesoscale structure of CMEs. Such relative arrangements are rare to attain fortuitously, but a few studies have taken advantage of a number of favourable periods featuring pairs of probes that happened to orbit close to each other in near-Earth space. For example, \citet{lugaz2018} and \citet{alalahti2020} analysed variations in the magnetic ejecta (for 21 events) and sheath region (for 29 events) structures, respectively, observed by spacecraft that were separated by ${\lesssim}$0.01~au in absolute distance. \citet{lugaz2018} found that, in the case of CME ejecta, the magnetic field magnitude has typical scale lengths of longitudinal magnetic coherence of 0.25--0.35~au (corresponding to 14--20$^{\circ}$ at 1~au), whilst the single magnetic components maintain their coherence over scales of 0.06--0.12~au (corresponding to 3--7$^{\circ}$ at 1~au). On the other hand, \citet{alalahti2020} found that, in the case of CME-driven sheaths, the largest scale of coherence is achieved by the east--west component of the magnetic field, with values lying around ${\sim}$0.15~au (corresponding to ${\sim}$9$^{\circ}$ at 1~au), whilst the magnetic field magnitude and the remaining components have typical scale lengths of 0.02--0.04~au (corresponding to 1--2$^{\circ}$ at 1~au). Other studies that investigated the magnetic configuration of CMEs over relatively short spatial scales at the heliocentric distance of ${\sim}$1~au include that of \citet{davies2021}, who analysed a single event and did not report significant differences in the structure observed at two spacecraft separated by ${\sim}$0.02~au radially and ${\sim}$1$^{\circ}$ longitudinally.

\begin{figure}[th!]
\centering
\includegraphics[width=\linewidth]{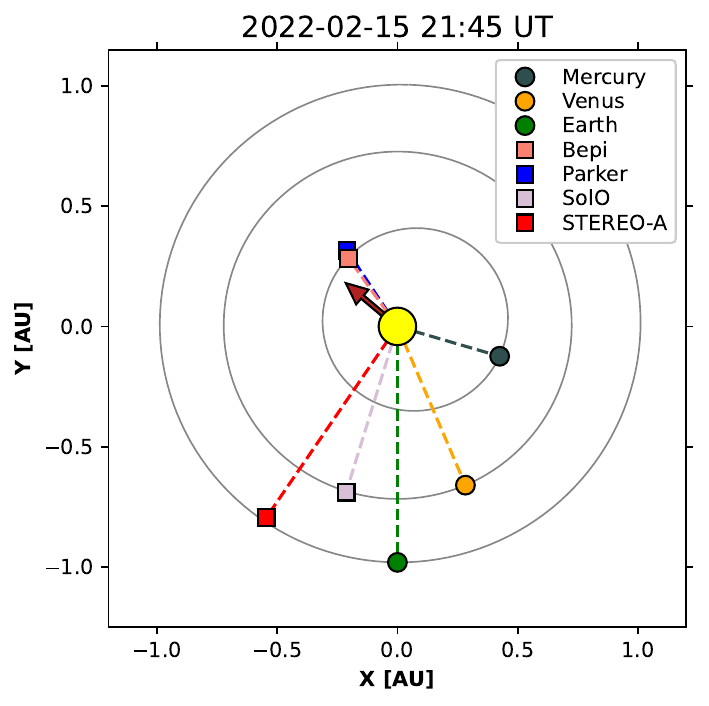}
\caption{Position of planets and spacecraft within 1~AU from the Sun on 2022 February 15 at 21:45~UT, i.e.\ around the CME eruption time. The propagation direction of the CME inferred from 3D reconstructions of the eruption in the corona is indicated with an arrow. The orbits of Mercury, Venus, and Earth are also shown.}
\label{fig:orbits}
\end{figure}

It is clear from the studies mentioned above that, in the context of CME research, the mesoscale region of the parameter space---defined by \citet{lugaz2018} as the critical region characterised by radial separations of 0.005--0.050~au and angular separations of of 1--12$^{\circ}$---is left largely unexplored to this day, especially considering that all investigations to date have been possible only via measurements collected at Earth's orbit. In this work, we analyse in detail the in-situ structure of the 2022 February 15 CME. This eruption has already gained attention in the heliophysics community because it represents the most-distant filament observed at extreme ultra-violet (EUV) wavelengths \citep{mierla2022}, but the scientific significance of this event goes beyond its exceptional remote-sensing measurements. In fact, the CME associated with this eruption was seen in situ by BepiColombo \citep[Bepi;][]{benkhoff2021} and Parker Solar Probe \citep[Parker;][]{fox2016}, which were relatively close to each other and only $\sim$0.35~au away from the Sun (see Figure~\ref{fig:orbits}). Hence, the 2022 February 15 CME represents the first occasion to investigate the mesoscale structure of CMEs at Mercury's orbit. The work presented here consists of a detailed analysis of the shock, sheath region, and magnetic ejecta observed at the two spacecraft, and aims to enhance our current understanding of the structure and evolution of CMEs at relatively early stages of their journey through interplanetary space.

\section{Overview of the 2022 February 15 Eruption} \label{sec:erupt}

In this section, we provide a brief overview of the remote-sensing observations associated with the 2022 February 15 eruption, in both the EUV and white-light (WL) regimes. For an in-depth discussion of the various sub-structures that can be identified in the available multi-spacecraft imagery, we direct the reader to \citet{mierla2022}.

\subsection{Extreme-ultra-violet Observations} \label{subsec:euv}

\begin{figure*}[th!]
\centering
\includegraphics[width=\linewidth]{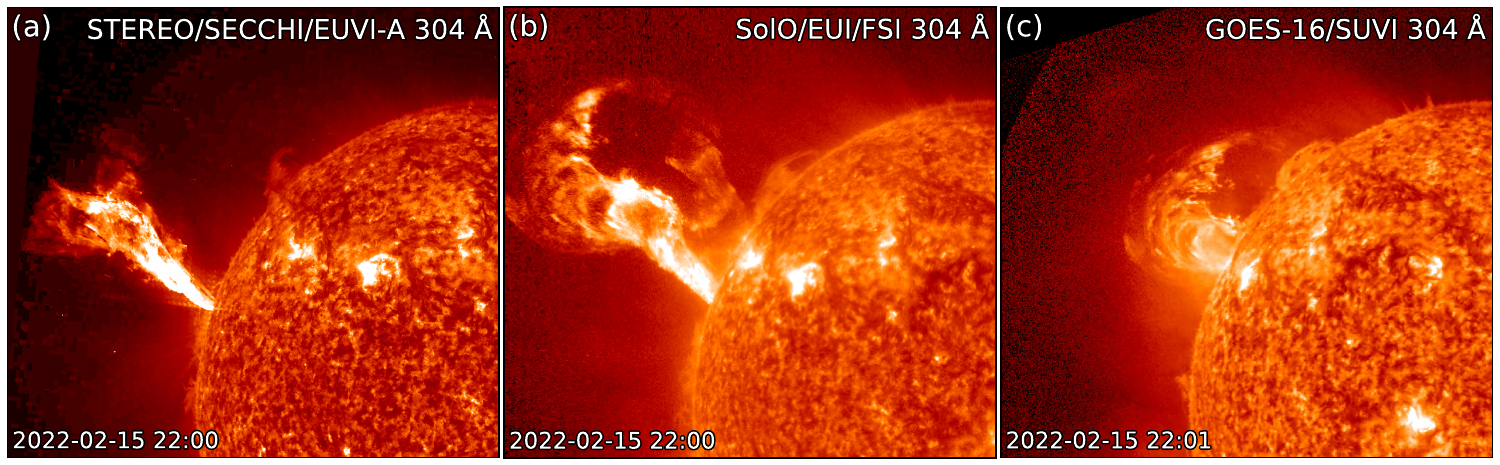}
\caption{The 2022 February 15 filament eruption at ${\sim}$22:00~UT as observed in EUV (304~{\AA} channel) from the available viewpoints. (a) STEREO/SECCHI/EUVI-A data. (b) SolO/EUI/FSI data. (c) GOES-16/SUVI data, representing Earth's view. In all images, the off-limb emission has been enhanced with a radial filter.}
\label{fig:euv_obs}
\end{figure*}

As mentioned in the Introduction, a large filament was involved in the 2022 February 15 eruption. The event was observed at EUV wavelengths from three viewpoints (see Figure~\ref{fig:orbits} for their positions), namely the Solar Terrestrial Relations Observatory Ahead \citep[STEREO-A;][]{kaiser2008} and Solar Orbiter \citep[SolO;][]{muller2020} spacecraft, as well as near-Earth---in this work, we use imagery from the Geostationary Operational Environmental Satellites 16 (GOES-16) probe. An overview of the remote-sensing EUV observations of the erupting filament from these three locations is shown in Figure~\ref{fig:euv_obs}, displaying data from the Extreme UltraViolet Imager (EUVI) telescope part of the Sun Earth Connection Coronal and Heliospheric Investigation \citep[SECCHI;][]{howard2008} suite onboard STEREO-A, the Full Sun Imager (FSI) telescope part of the Extreme Ultraviolet Imager \citep[EUI;][]{rochus2020} suite onboard SolO, and the Solar UltraViolet Imager \citep[SUVI;][]{darnel2022} telescope onboard GOES-16. From these observations, it is clear that the 2022 February 15 eruption appeared as a behind-the-eastern-limb event from all three viewpoints, albeit to different extents. If we consider the average of the spatial range spanned by various filament and CME features in 3D as triangulated by \citet{mierla2022}, i.e.\ E138N35 in Stonyhurst coordinates, then the event took place behind the eastern limb by ${\sim}$15$^{\circ}$ for STEREO-A, ${\sim}$30$^{\circ}$ for SolO, and ${\sim}$50$^{\circ}$ for Earth.

The erupting filament was first observed to emerge from behind the northeastern limb from all three viewpoints around 21:50~UT on 2022 February 15. Its fine structure was best visible at 304~{\AA} (i.e., the channel shown in Figure~\ref{fig:euv_obs}), but its appearance was prominent also at 171~{\AA} from all three imagers, in addition to being observed at several other EUV wavelengths (depending on the availability of each instrument). During the early phases of the eruption, the filament appeared to be surrounded by a ``bubble''-like leading edge \citep[see Figures~2 and 5 in][]{mierla2022}, the resulting structure being reminiscent of the classic three-part CME that is usually observed in coronagraph data \citep[e.g.,][]{illing1985, vourlidas2013}. Throughout its evolution across the fields of view of the three different telescopes \citep[we note that the SolO/EUI/FSI field extended up to $>$6\,$R_{\odot}$ at the time of this event, see][]{mierla2022}, the filament propagated in a northeasterly direction, away from the solar equatorial plane. This suggests that any near-the-ecliptic probe detecting this CME in situ would likely experience a flank encounter rather than a central one, assuming that no dramatic equatorward deflections would take place in the outer corona and/or interplanetary space \citep[since stronger CMEs tend to be less affected by the configuration of the coronal and interplanetary magnetic field; e.g.,][]{gui2011, kay2015}.

\subsection{White-light Observations} \label{subsec:wl}

\begin{figure}[t!]
\centering
\includegraphics[width=\linewidth]{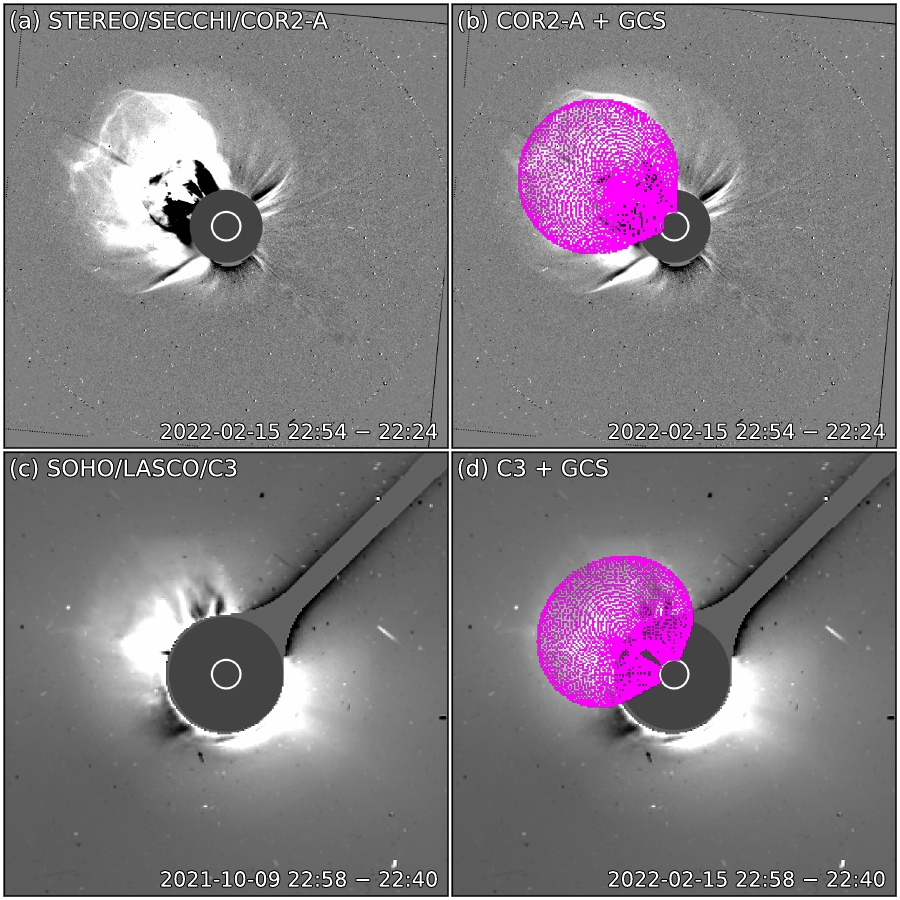}
\caption{The 2022 February 15 CME at ${\sim}$23:00~UT as observed in WL (coronagraph) difference imagery from the available viewpoints. (a) STEREO/SECCHI/COR2-A data, shown also in (b) with the GCS wireframe overlaid. (c) SOHO/LASCO/C3 data, shown also in (d) with the GCS wireframe overlaid.}
\label{fig:wl_obs}
\end{figure}

The 2022 February 15 eruption was observed in WL imagery from two viewpoints, i.e.\ the STEREO-A spacecraft and Earth---although SolO is equipped with a coronagraph, the instrument was not operational at the time of this event. From the STEREO-A viewpoint, the apex of the CME first appeared around 21:55~UT in the COR1 field of view and around 22:10~UT in the COR2 one, both cameras being part of the SECCHI suite. From Earth's viewpoint, the CME was observed by the Large Angle Spectroscopic Coronagraph \citep[LASCO;][]{brueckner1995} onboard the Solar and Heliospheric Observatory \citep[SOHO;][]{domingo1995}, emerging around 22:10~UT in the C2 field of view and around 22:30~UT in the C3 one. An overview of the remote-sensing WL observations of the CME in the solar corona from these two locations is shown in Figure~\ref{fig:wl_obs}.

From coronagraph data, it is evident that the 2022 February 15 CME displayed clear shock signatures, usually manifested as a fainter emission that surrounds the brighter magnetic ejecta \citep[e.g.,][]{vourlidas2003, kouloumvakos2022}. Additionally, the CME propagated towards the northeast from both available viewpoints, in agreement with the EUV observations presented in Section~\ref{subsec:euv}. To obtain a more quantitative estimate of the CME geometric and kinematic parameters, we fit the Graduated Cylindrical Shell \citep[GCS;][]{thernisien2011} model to WL data. The GCS model consists of a parameterised shell (described by six variables) that can be projected onto nearly-simultaneous coronagraph images from different perspectives and is then manually adjusted to best match the CME structures seen in the data. Despite being associated with uncertainties due to the user's subjectivity when performing a fit \citep[e.g.,][]{verbeke2023}, the GCS model is widely used in CME and space weather research \citep[see, e.g.,][for some recent applications]{nieveschinchilla2022, palmerio2022a, rodriguezgarcia2022}. An example of GCS fitting applied to the 2022 February 15 CME is shown in the right panels of Figure~\ref{fig:wl_obs}. According to our results, the CME apex propagates at a latitude of 33$^{\circ}$ and a longitude of $-$130$^{\circ}$ (both values are given in Stonyhurst coordinates), the CME axis has a tilt of $-$60$^{\circ}$ with respect to the solar equator (defined positive for counterclockwise rotations from the solar west direction), the CME half-angular width along its major axis is ${\sim}$45$^{\circ}$, and the CME radial speed in the solar corona is ${\sim}$2200~km$\cdot$s$^{-1}$. These values are in excellent agreement with those reported in \citet{mierla2022}, who estimated a propagation direction of 34$^{\circ}$ in latitude and $-$132$^{\circ}$ in longitude, as well as a speed of ${\sim}$2200~km$\cdot$s$^{-1}$. It is evident from the GCS fitting shown in Figure~\ref{fig:wl_obs} and our resulting parameters that the 2022 February 15 CME was characterised by a high latitude and a high inclination. This indicates that a near-the-ecliptic spacecraft encountering the ejecta in situ would likely cross its southern flank, as already speculated when considering EUV observations (see Section~\ref{subsec:euv}).

After leaving the solar corona, the 2022 February 15 CME was observed in WL heliospheric imagery by the Wide-Field Imager for Solar Probe Plus \citep[WISPR;][]{vourlidas2016} suite of cameras onboard Parker. Only the southern portion of the full structure was captured in the fields of view of the WISPR telescopes, further proving that the CME continued propagating in interplanetary space with a prominent northward component, and that equatorward deflections are not expected to have taken place. These observations are presented and briefly discussed in Appendix~\ref{app:wispr}.


\section{In-situ Observations at Bepi and Parker} \label{sec:insitu}

In this section, we provide an overview of the available in-situ measurements of the 2022 February 15 CME at Bepi and Parker. We first show a comparison of the magnetic field data at both spacecraft, followed by a more detailed summary of magnetic field and plasma observations at Parker---the plasma instruments onboard Bepi are not regularly operational during cruise phase (i.e., preceding the Mercury orbit insertion scheduled for December 2025). Additionally, we remark that the magnetic field data from Bepi have been subject to additional processing due to the spacecraft's solar electric propulsion system (SEPS) being activated during the period of interest (more information is provided in Appendix~\ref{app:mpomag}). \edit1{The timings of the various CME-related features discussed in the following sections, together with the exact positions of Bepi and Parker when crossing each, are summarised in Table~\ref{tab:positions}.}

\begin{table}[t!]
\caption{Times and positions of Bepi and Parker when crossing CME-related features. \label{tab:positions}}
\centering
\renewcommand{\arraystretch}{1}
\begin{tabularx}{\linewidth}{l@{\hskip .25in}cc}
\toprule
 & \textsc{\textbf{Time}} & \textsc{\textbf{Position}} \\
  & [UT] & [r, ${\theta}$, ${\phi}$] \\
\midrule
\textsc{\textbf{Bepi}}\\  
\cmidrule(r{10pt}){1-1}
Shock & 2022-02-16T06:25 & [0.35, 1.71, $-142.73$] \\
Ejecta start & 2022-02-16T15:41 & [0.34, 1.78, $-140.99$] \\
Ejecta end & 2022-02-17T07:51 & [0.34, 1.91, $-137.89$] \\
\midrule
\textsc{\textbf{Parker}}\\  
\cmidrule(r{10pt}){1-1}
Shock & 2022-02-16T07:25 & [0.37, 3.71, $-145.92$] \\
Cloud start & 2022-02-16T15:18 & [0.36, 3.70, $-145.43$] \\
Cloud end & 2022-02-17T06:32 & [0.35, 3.67, $-144.39$] \\
Obstacle end & 2022-02-17T16:48 & [0.33, 3.64, $-143.58$] \\
\bottomrule
\end{tabularx}
\vspace*{.1in}
\begin{tablenotes}
\item \emph{Notes.} The [r, ${\theta}$, ${\phi}$] triplets are reported in Stonyhurst coordinates in units of [au, $^{\circ}$, $^{\circ}$]. See the text throughout Section~\ref{sec:insitu} for a discussion on the identification of all the boundaries.
\end{tablenotes}
\end{table}

\subsection{Magnetic Field Measurements at the Two Spacecraft} \label{subsec:mag}

\begin{figure*}[th!]
\centering
\includegraphics[width=\linewidth]{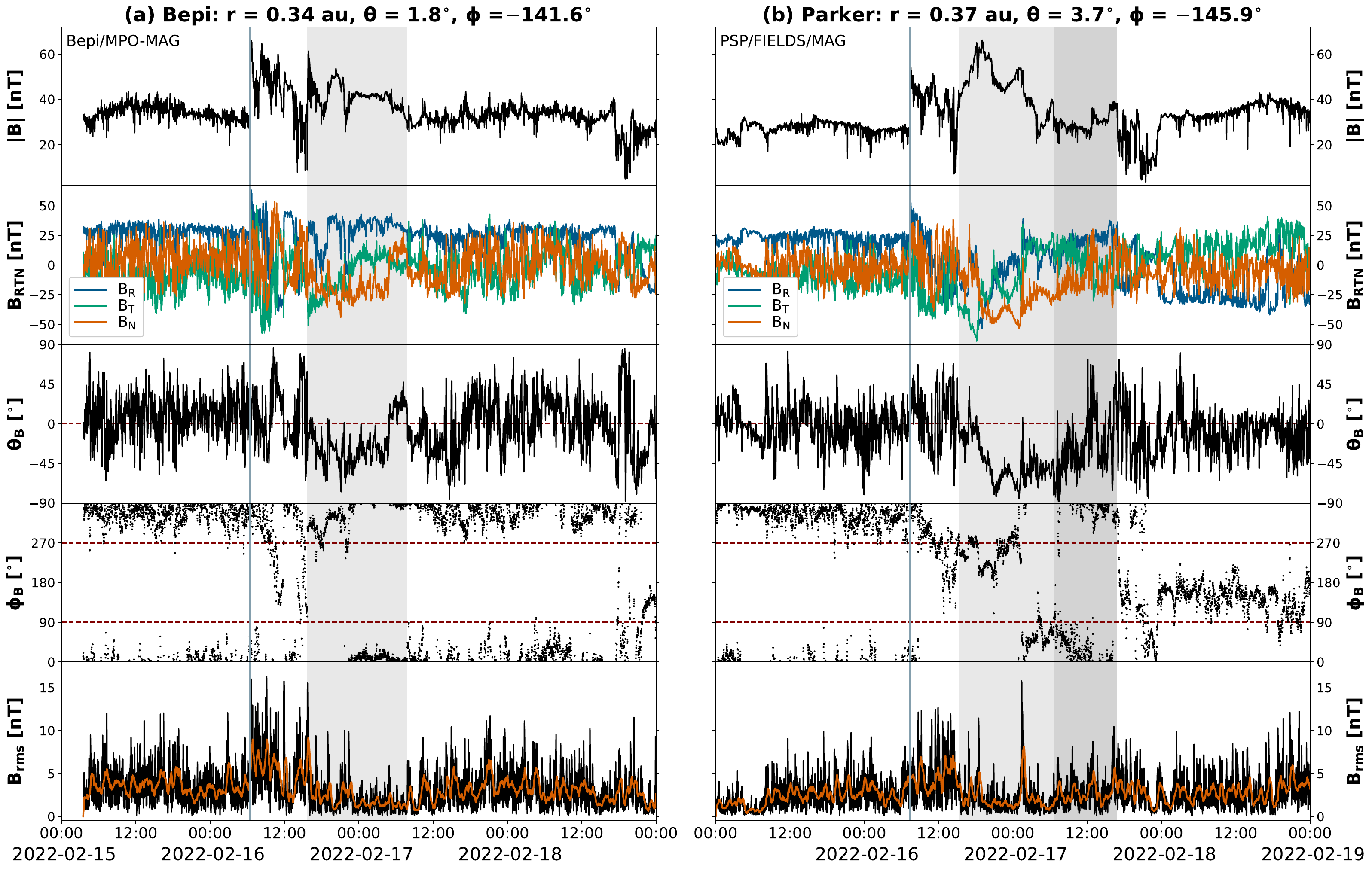}
\caption{Magnetic field measurements (1-min cadence) at (a) Bepi and (b) Parker for the period 2022 February 15--18. Each plot shows, from top to bottom: magnetic field magnitude, magnetic field Cartesian components in radial--tangential--normal (RTN) coordinates, latitudinal angle of the magnetic field, longitudinal angle of the magnetic field, and root-mean-square error of the magnetic field (using a rolling-average over 15~min for each component, the superposed orange curve displays the 30-min rolling average of the same data). The ($r$, $\theta$, $\phi$) values displayed at the top of the panels for both spacecraft refer to 2022 Feburary 16, 07:00~UT, and the angles are reported in Stonyhurst coordinates. The vertical grey lines indicate the arrival of the CME-driven shock at either spacecraft, whilst the shaded grey areas highlight the CME ejecta. At Parker, the full ejecta (magnetic obstacle) includes both grey regions, and the flux rope (magnetic cloud) structure within it is represented by the lighter grey area. Additional details about the identification of boundaries at Parker are provided in Section~\ref{subsec:psp}.}
\label{fig:insitu_mag}
\end{figure*}

Given the high speed of the 2022 February 15 CME as inferred from remote-sensing observations (see Section~\ref{subsec:wl}), we can expect a relatively rapid propagation to ${\sim}$0.35~au, i.e., the heliocentric distance of Bepi and Parker at the time of the event (see Figure~\ref{fig:orbits}). The two probes were located ${\sim}$15$^{\circ}$ to the east and ${\sim}$30$^{\circ}$ to the south of the CME nose trajectory that was estimated from GCS fitting. By inspecting magnetic field measurements from the Mercury Planetary Orbiter Magnetometer \citep[MPO-MAG;][]{heyner2021} onboard Bepi and the Fluxgate Magnetometer (MAG) part of the FIELDS \citep{bale2016} investigation onboard Parker---both data sets are shown in Figure~\ref{fig:insitu_mag}---we find signatures of a solar wind transient starting on February 16 around 07:00~UT (approximate average time between the two spacecraft). This yields a travel time of ${\sim}$9.25~h, with an average transit speed of \edit1{1620}~km$\cdot$s$^{-1}$. Specifically, we observe an abrupt increase in the magnetic field magnitude at 06:25~UT in Bepi data and at 07:25~UT in Parker data (marked with vertical grey lines in Figure~\ref{fig:insitu_mag}), indicating the passage of a CME-driven shock\edit1{---by considering the exact shock arrival times and positions for each spacecraft, we obtain average transit speeds of 1672~km$\cdot$s$^{-1}$ at Bepi and 1606~km$\cdot$s$^{-1}$ at Parker}. The identification of this feature as a fast-forward shock can be confirmed using plasma data from Parker (presented in detail in Section~\ref{subsec:psp}), hence it is reasonable to assume that the same holds true for Bepi.

After a period (${\sim}$8.5~h) characterised by highly fluctuating magnetic field vectors, corresponding to the turbulent sheath region that usually follows CME-driven shocks, \edit1{around 15:30~UT (approximate average time between the two spacecraft)} both probes observed enhanced magnetic field magnitudes, together with smoother profiles of the field components, which can be attributed to the passage of the CME ejecta (highlighted as shaded grey areas in Figure~\ref{fig:insitu_mag}). \edit1{This corresponds to a travel time of ${\sim}$17.67~h, with an average transit speed of \edit1{823}~km$\cdot$s$^{-1}$.} Specifically, the leading edge of this structure is identified on February 16 at 15:41~UT for Bepi and 15:18~UT for Parker\edit1{, resulting in average ejecta transit speeds of 798~km$\cdot$s$^{-1}$ and 861~km$\cdot$s$^{-1}$, respectively. We note that Parker observed the ejecta leading edge slightly earlier despite being ${\sim}0.02$~au farther from the Sun than Bepi at the time (see Table~\ref{tab:positions}). This suggests that Parker may have been somewhat closer to the CME nose given its higher heliocentric latitude and the CME propagation direction being mainly above the ecliptic, or that the probes encountered a highly distorted CME front.} At both spacecraft, the magnetic field vectors rotated smoothly from east to west and pointed mostly southwards throughout the structure, suggesting approximately an east--south--west (ESW, right-handed) flux rope type from visual inspection \citep[e.g.,][]{bothmer1998, palmerio2018}. Whilst at Bepi the entire identified CME ejecta interval was characterised by flux-rope-like signatures, at Parker the trailing portion of the structure featured more turbulent magnetic fields (especially in the north--south component), suggesting that the flux rope and, more generally, the ejecta boundaries may not coincide \citep[e.g.,][more details on this matter are provided in Section~\ref{subsec:psp}]{richardson2010, kilpua2013}. For simplicity and to avoid confusion, we shall refer henceforth to the flux rope interval as `magnetic cloud' \citep[e.g.,][]{burlaga1981} and to the whole CME ejecta as `magnetic obstacle' \citep[e.g.,][]{nieveschinchilla2019}.

We also investigate the level of fluctuations in the interplanetary magnetic field by computing the root-mean-square of the field vector, shown in Figure~\ref{fig:insitu_mag}, using


\begin{equation}
B_\mathrm{rms} = \sqrt{ \sum_{i=R,T,N} \frac{( B_{i} - {\langle}B_{i}{\rangle} )^{2}}{n} } \, , 
    \label{eq:rms}
\end{equation}

\noindent where for the averaged quantities we employ 15-min rolling averages of the corresponding magnetic field component and $n$ is the number of data points employed for each average (hence, $n=15$ here). It is evident that the portion of solar wind identified as the sheath region in either data set features generally the most fluctuations when compared to the ambient medium and to the CME ejecta material, in agreement with statistical studies of interplanetary CMEs detected at 1~au \citep[e.g.,][]{masiasmeza2016, kilpua2019b}. At Bepi, the level of fluctuations displays a rather clear decrease inside the magnetic ejecta with respect to the entire time series, as expected from statistical studies of CMEs measured at 1~au \citep[e.g.,][]{rodriguez2016, regnault2020}. At Parker, the same holds generally true, but we note that the largest fluctuations of the entire time series are reached briefly in the middle of the ejecta due to a sudden change in the magnetic field vector. Additionally, we observe a relatively higher level of fluctuations in the second part of the ejecta (i.e., outside of the magnetic cloud interval), which is to be expected since flux rope ejecta tend to display more organised magnetic fields \citep[e.g.,][]{li2018, nieveschinchilla2019}.

\subsection{Magnetic Field and Plasma Measurements at Parker} \label{subsec:psp}

\begin{figure*}[p!]
\centering
\includegraphics[width=.88\linewidth]{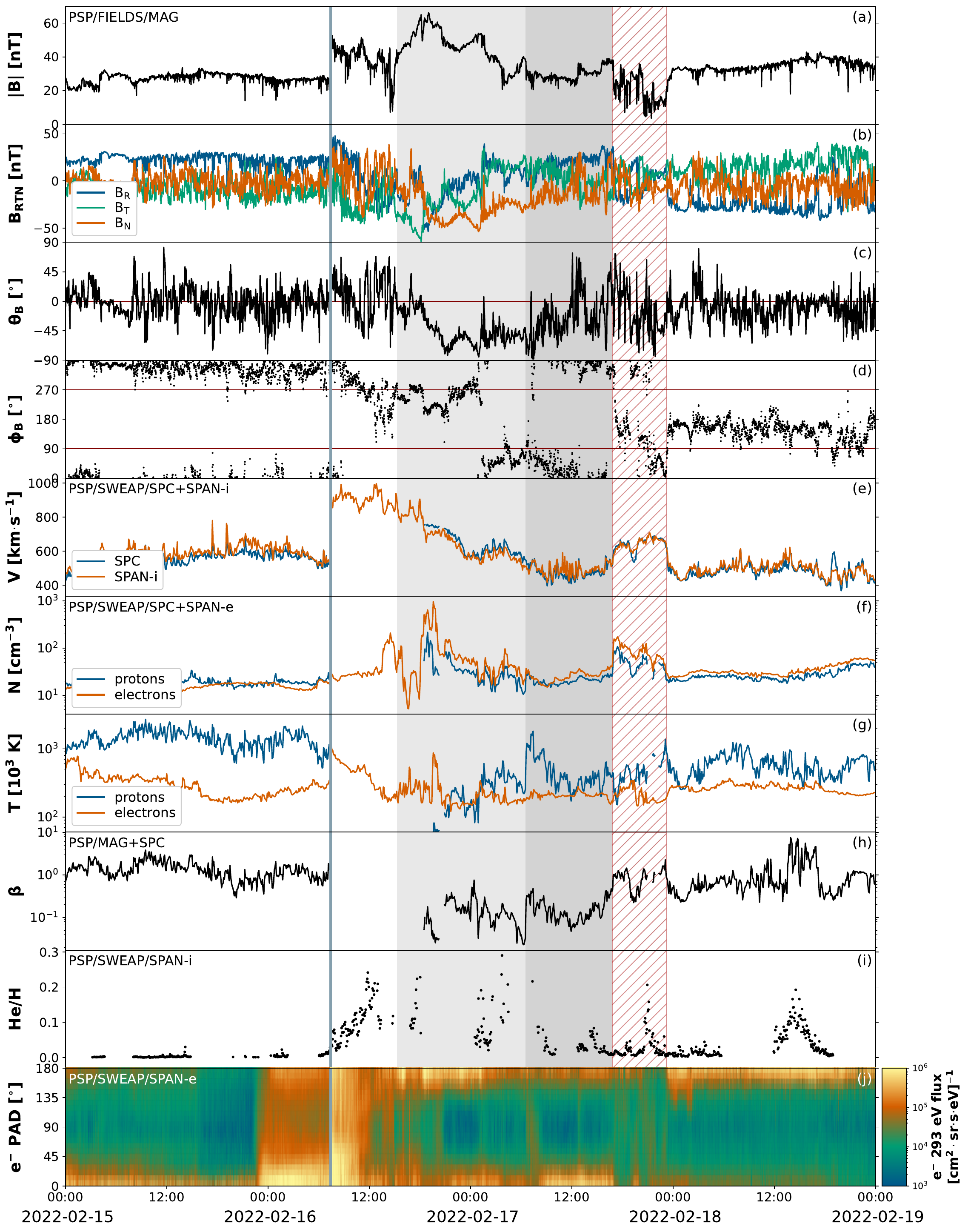}
\caption{Overview of in-situ data at Parker for the period 2022 February 15--18. The plot shows, from top to bottom: (a) magnetic field magnitude, (b) magnetic field Cartesian components in RTN coordinates, (c) latitudinal and (d) longitudinal angles of the magnetic field, (e) solar wind bulk speed, proton (f) density and (g) temperature, (h) plasma beta, (i) helium-to-proton ratio, and (j) pitch-angle spectrogram of suprathermal 293~eV electrons. The vertical grey line indicates the arrival of the CME-driven shock, whilst the shaded grey areas highlight the magnetic obstacle (light+dark grey) and the magnetic cloud (light grey) within it. The hatched area marks the HPS passage immediately following the CME.}
\label{fig:insitu_psp}
\end{figure*}

As mentioned at the beginning \edit1{of} this section, the available in-situ observations at Parker for the 2022 February 15 event include plasma data, which, on the other hand, do not find a direct comparison at Bepi due to the corresponding instrument(s) being not operational. The whole set of in-situ observations at Parker is presented in Figure~\ref{fig:insitu_psp}, where the FIELDS magnetic field data are complemented with plasma measurements from the Solar Wind Electrons Alphas and Protons \citep[SWEAP;][]{kasper2016} suite---specifically, using the Solar Probe Cup \citep[SPC;][]{case2020}, Solar Probe ANalyzer-Ions \citep[SPAN-i;][]{livi2022}, and Solar Probe ANalyzer-Electrons \citep[SPAN-e;][]{whittlesey2020} detectors. Here, we use the collective of these observations to better constrain and define the various CME features and sub-structures that were briefly introduced in Section~\ref{subsec:mag}. We note that there is a data gap in the SPC proton moments immediately following the detection of the CME-driven shock due to the solar wind being outside of the instrument's energy range (bulk speed ${\gtrsim}$800~km$\cdot$s$^{-1}$). SPAN-i data, on the other hand, despite capturing the profile of the solar wind speed in its entirety, considerably underestimates the proton density and temperature due to the full ion distribution not being in the instrument's field of view. Hence, we complement the available proton density and temperature measurements from SPC with electron data from SPAN-e derived from fitting to the electron core distribution \citep[using the procedure described in][]{romeo2023}.

First of all, we note that the abrupt magnetic field jump that we had identified as the CME-driven shock in Section~\ref{subsec:mag} (vertical grey line in Figure~\ref{fig:insitu_psp}) is indeed accompanied by a prominent rise in the solar wind bulk speed (from ${\sim}$550 to ${\sim}$850~km$\cdot$s$^{-1}$), a moderate climb in the electron density and temperature, as well as an intensification of suprathermal electron fluxes \edit1{\citep[the electron flux enhancements observed prior to the shock arrival are due to the strong particle event associated with this eruption, analysed e.g.\ by][]{giacalone2023, khoo2024}}. The following sheath region is characterised by highly fluctuating magnetic fields, almost-constant speeds, a declining temperature profile, and a plateauing density trend except for a high-density feature close to its trailing portion. The density peak at the back of the sheath is reminiscent of a piled-up compression (PUC) region that forms ahead of the nose of a driver (in this case, the CME ejecta) often due to its large expansion speed, which is known to usually be higher close to the Sun and then to decrease gradually as the CME travels through interplanetary space \citep[e.g.,][]{farrugia2008, das2011}.

Regarding the CME ejecta (shaded grey area in Figure~\ref{fig:insitu_psp}), we use the plasma measurements to corroborate our initial assessment of non-coinciding magnetic cloud and magnetic obstacle boundaries (see Section~\ref{subsec:mag}). Indeed, we note that bidirectional suprathermal electrons---i.e., with the pitch angle distribution (PAD) displaying peaks close to 0$^{\circ}$ and 180$^{\circ}$---are detected throughout the identified CME ejecta interval. This feature is usually interpreted as the result of two electron beams flowing along and against the magnetic field lines, and are often associated with closed lines that are connected to the Sun from both ends, especially in the case of CME ejecta \citep[e.g.,][]{gosling1987, carcaboso2020}. Additionally, we find that the magnetic cloud period (characterised by smoother and rotating magnetic fields) features a declining speed profile, usually interpreted as a signature of expansion \citep[e.g.,][]{zurbuchen2006, nieveschinchilla2018a}. The trailing portion of the magnetic obstacle (where magnetic fluctuations are higher), on the other hand, displays first a plateauing and then a slightly rising speed trend, suggesting that expansion of the structure had likely ceased in that region and possibly related to the small rise in magnetic field magnitude over the last ${\sim}$5~hr of the ejecta passage. The temperature and plasma beta show rather irregular profiles, but they display periods with values below the ambient ones throughout the ejecta interval. The density is generally never below ambient values, and features an isolated peak inside the magnetic cloud interval (coinciding with a local maximum in the electron temperature as well). A possible cause for the observed mismatching magnetic cloud and magnetic obstacle boundaries is flux erosion (more details on this aspect are presented in Section~\ref{subsec:ejecta}). 

From the He/H composition panel, we note that the He abundances rise throughout the sheath towards the ejecta (grey region in Figure~\ref{fig:insitu_psp}). Prior to the shock, the solar wind He/H abundance appears below 1\%, which is typical of solar wind values \citep[e.g.,][]{borrini1981}. The gradual rise in the He/H values throughout the sheath indicates there is no sharp boundary between the solar wind, the sheath/shocked plasma, and the main CME ejecta. The smooth transition is consistent with a gradual accumulation and mixture of solar wind and CME material in the front. Moreover, there are also some He/H enhancements that appear throughout the trailing part of the ejecta, which is in qualitative agreement with the \citet{rodriguez2016} superposed epoch profile obtained from 1~au events and may result from He gravitational settling in the pre-eruption CME source region \citep[e.g.,][]{richardson2004}. The fluctuations in the He abundance could indicate the passage of coherent substructure in connection to the filament's fragmented nature observed in the remote-sensing images (see Section~\ref{subsec:euv}). However, due to the field of view restrictions of SPAN-i, many of the measurements are omitted across this time frame, making it difficult to examine with certainty. Previous work has shown that He/H abundances are generally elevated in CMEs compared to the solar wind \citep{borrini1982, zurbuchen2016}. However, it remains unclear whether filaments themselves contain high He/H abundances \citep{lepri2021, rivera2022}. For a similarly energetic CME, the Bastille Day CME that erupted on 2000 July 14 and reached 2000~km$\cdot$s$^{-1}$ in the low corona, its compositional structure indicated that the prominence material had elemental abundances that were less enhanced in the low first-ionization-potential species than the surrounding flux rope structure, suggesting more solar wind-like elemental properties \citep{rivera2023}.

Finally, we note a structure immediately following the ejecta characterised by a low magnetic field magnitude as well as enhanced speed, density, and plasma beta (hatched area in Figure~\ref{fig:insitu_psp}). Additionally, both the longitudinal component of the magnetic field and the suprathermal electron PAD display a ${\sim}$180$^{\circ}$ reversal after the passage of this feature, compared to the pre-CME ambient wind. The structure shows all the typical characteristics of a heliospheric plasma sheet (HPS) crossing \citep[e.g.,][]{crooker2004, lavraud2020}, i.e., the passage of a high-density and high-beta region that surrounds the heliospheric current sheet (HCS). We tentatively associate the HPS at Parker (starting at ${\sim}$17:00~UT on February 17) with a similar structure (featuring a low magnetic field magnitude as well as a reversal of the longitudinal component of the field) detected at Bepi approximately one day later (starting at ${\sim}$17:00~UT on February 18, see Figure~\ref{fig:insitu_mag}), i.e.\ well after the passage of the 2022 February 15 CME. Hence, the (spatial and temporal) proximity of the HPS to the CME at Parker---and the resulting interaction between the two---may explain the presence of more ambiguous magnetic cloud and magnetic obstacle boundaries at Parker, which are on the other hand not found at Bepi.


\section{Analysis of the Mesoscale CME Structure} \label{sec:analysis}

In this section, we analyse in detail the similarities and differences related to observations of the 2022 February 15 CME at Bepi and Parker. We focus our investigation on the three main regions and/or features of in-situ CMEs, i.e., the interplanetary shock, the sheath region, and the magnetic ejecta. 

\subsection{Interplanetary Shock} \label{subsec:shock}

As mentioned in Section~\ref{subsec:mag}, an interplanetary shock was detected on 2022 February 16 at 06:25~UT by Bepi and 07:25~UT by Parker. Although it is not possible to determine the passage of a shock with certainty at Bepi due to the lack of corresponding plasma data, the prominent nature of the shock at Parker (see Section~\ref{subsec:psp}), together with the spatial proximity of the two probes, allows us to proceed with our analysis under the assumption that both spacecraft measured the same feature.

\begin{table*}[t!]
\caption{Shock parameters derived for both Bepi and Parker. \label{tab:shock} \vspace*{.1in}}
\centering
\renewcommand{\arraystretch}{1}
\begin{tabular}{l@{\hskip 1in}c@{\hskip .7in}c}
\toprule
 & \textsc{\textbf{BepiColombo}} & \textsc{\textbf{Parker Solar Probe}} \\
\midrule
\textsc{\textbf{Jumps}}\\  
\cmidrule(r{50pt}){1-1}
$r_B$ & 2.04 & 2.04 \\
$r_n$ & --- & 1.80 \\
${\Delta}V$ & --- & 282 km$\cdot$s$^{-1}$ \\
\midrule
\textsc{\textbf{MCT}}\\  
\cmidrule(r{50pt}){1-1}
$\mathbf{\hat{n}}_\mathrm{MCT}$ & [0.76, 0.63, 0.16] & [0.45, $-$0.82, $-$0.37] \\
$\Theta_\mathrm{MCT}$ & 7.3$^{\circ}$ & 31.8$^{\circ}$ \\
$\Lambda_\mathrm{MCT}$ & 40.5$^{\circ}$ & 63.3$^{\circ}$ \\
$\Xi_\mathrm{MCT}$ & 9.0$^{\circ}$ & $-$21.4$^{\circ}$ \\
\midrule
\textsc{\textbf{MVA}}\\  
\cmidrule(r{50pt}){1-1}
$\mathbf{\hat{n}}_\mathrm{MVA}$ & [0.57, $-$0.25, $-$0.78] & [0.53, $-$0.42, $-$0.73] \\
$\Theta_\mathrm{MVA}$ & 80.4$^{\circ}$ & 42.6$^{\circ}$ \\
$\Lambda_\mathrm{MVA}$ & 55.0$^{\circ}$ & 57.8$^{\circ}$ \\
$\Xi_\mathrm{MVA}$ & $-$51.3$^{\circ}$ & $-$47.1$^{\circ}$ \\
$\lambda_{2}$/$\lambda_{3}$ & 3.07 & 2.71 \\
$B_{n}$/$B$ & 0.15 & 0.45 \\
\midrule
\textsc{\textbf{MD3}}\\  
\cmidrule(r{50pt}){1-1}
$\mathbf{\hat{n}}_\mathrm{MD3}$ & --- & [0.49, $-$0.71, $-$0.51] \\
$\Theta_\mathrm{MD3}$ & --- & 33.3$^{\circ}$ \\
$\Lambda_\mathrm{MD3}$ & --- & 60.9$^{\circ}$ \\
$\Xi_\mathrm{MD3}$ & --- & $-$30.6$^{\circ}$ \\
\bottomrule
\end{tabular}
\vspace*{.1in}
\begin{tablenotes}
\item \emph{Notes.} $r_B$ = magnetic compression ratio. $r_n$ = density compression ratio. ${\Delta}V$ = velocity jump. $\mathbf{\hat{n}}$ = shock normal. $\Theta$ = shock angle. $\Lambda$ = location angle. $\Xi$ = elevation angle. $\lambda_{2}$/$\lambda_{3}$ = intermediate-to-minimum eigenvalue ratio. $B_{n}$/$B$ = minimum-variance-to-total magnetic field ratio. MCT = magnetic coplanarity theorem. MVA = minimum variance analysis. MD3 = mixed-mode method.
\end{tablenotes}
\end{table*}

When investigating the properties of interplanetary shocks, the first step is to define the plasma states upstream and downstream of the shock. In this work, the criteria for determining the upstream and downstream conditions are adopted from those used in the Heliospheric Shock Database \citep[HSD\footnote{The Heliospheric Shock Database is generated and maintained at the University of Helsinki and can be accessed at \href{http://ipshocks.fi}{http://ipshocks.fi}};][]{kilpua2015}, and consist of averaging the quantities under consideration over 8~min intervals that end 1~min before and start 2~min after the identified shock passage time, i.e.:

\begin{equation}
    \begin{cases}
      t_{u} = [\, t_\mathrm{shock} - 9 \mathrm{min}, t_\mathrm{shock} - 1 \mathrm{min} \,]\\
      t_{d} = [\, t_\mathrm{shock} + 2 \mathrm{min}, t_\mathrm{shock} + 10 \mathrm{min} \,] \,,
    \end{cases}
\label{eq:updown}
\end{equation}

\noindent where the subscript `\emph{u}' indicates the upstream state and the subscript `\emph{d}' indicates the downstream state. These intervals ensure that the mean upstream and downstream values are estimated sufficiently far from the shock ramp (and the particularly turbulent region in the immediate shock downstream). We remark, nevertheless, that the choice of the upstream and downstream windows may significantly affect the resulting shock parameters \citep[e.g.,][]{balogh1995, trotta2022}. Here, small variations in the interval selections had minimal impact on our calculated normals, but we did not perform an exhaustive parameter-dependency analysis. The various shock properties calculated for both Bepi (where possible) and Parker are summarised in Table~\ref{tab:shock}, and they are described in detail below.

The first two shock characteristics reported in Table~\ref{tab:shock} are compression ratios, defined as

\begin{equation}
    r_X = \frac{X_d}{X_u} \, ,
\label{eq:compression}
\end{equation}

\noindent where $X$ indicates a general physical quantity---usually the magnetic field magnitude, $B$, the proton density, $n_{p}$, and/or the proton temperature, $T_{p}$. In the case of the event studied here, we report the magnetic compression ratio, $r_B$, for both Bepi and Parker, whilst for the density compression ratio, $r_n$, we use electron measurements at Parker as a proxy---generally, a reasonable approximation under the assumption of quasi-neutrality of the plasma---due to the lack of downstream proton data (see Figure~\ref{fig:insitu_psp}). We note that $r_B$ features identical values of 2.04 at the two spacecraft, whilst $r_n$ is found to equal 1.80 at Parker. These compression ratios are usually considered as proxies for the strength of a shock \citep[e.g.,][]{burton1996, oh2007}, and their expected values have been found to not vary significantly between 0.3 and 1~au \citep{volkmer1985, lai2012}. We find our compression ratios to be very similar to the median values detected by the Helios mission between 0.3 and 0.8~au---i.e., 1.84 for $r_B$ and 1.87 for $r_n$ \citep{perezalanis2023}, indicating that the 2022 February 15 CME drove a shock of average strength. Nevertheless, studies have found that interplanetary shocks tend to be stronger in the proximity of their nose than along their flanks \citep[e.g.,][]{cane1988, kallenrode1993}, suggesting that had the CME been encountered closer to its apex (see Section~\ref{subsec:wl} for an estimate of its propagation direction well north of the ecliptic), Bepi and Parker may have measured higher values for the corresponding compression ratios.

The next parameter reported in Table~\ref{tab:shock} is the shock speed jump, defined as

\begin{equation}
    {\Delta}V = V_d - V_u \, .
\label{eq:jump}
\end{equation}

As in the case of $r_n$, ${\Delta}V$ can be calculated at Parker only, resulting in a jump of 282~km$\cdot$s$^{-1}$. Although shocks driven by fast CMEs tend to slow down as they propagate through the inner solar system \citep[e.g.,][]{woo1985, neugebauer2013}, this value is slightly higher than average, since shock jumps measured inside 1~au are usually below 200~km$\cdot$s$^{-1}$ \citep{luhmann1995}. The high speed of the shock driven by the 2022 February 15 CME and detected at ${\sim}$0.35~au may be due to its propagation through a rarefied (${\sim}$20~cm$^{-3}$) and fast (${\sim}$600~km$\cdot$s$^{-1}$) solar wind stream (see Figure~\ref{fig:insitu_psp}), which in turn hindered its deceleration due to drag effects.

\begin{figure*}[th!]
\centering
\includegraphics[width=\linewidth]{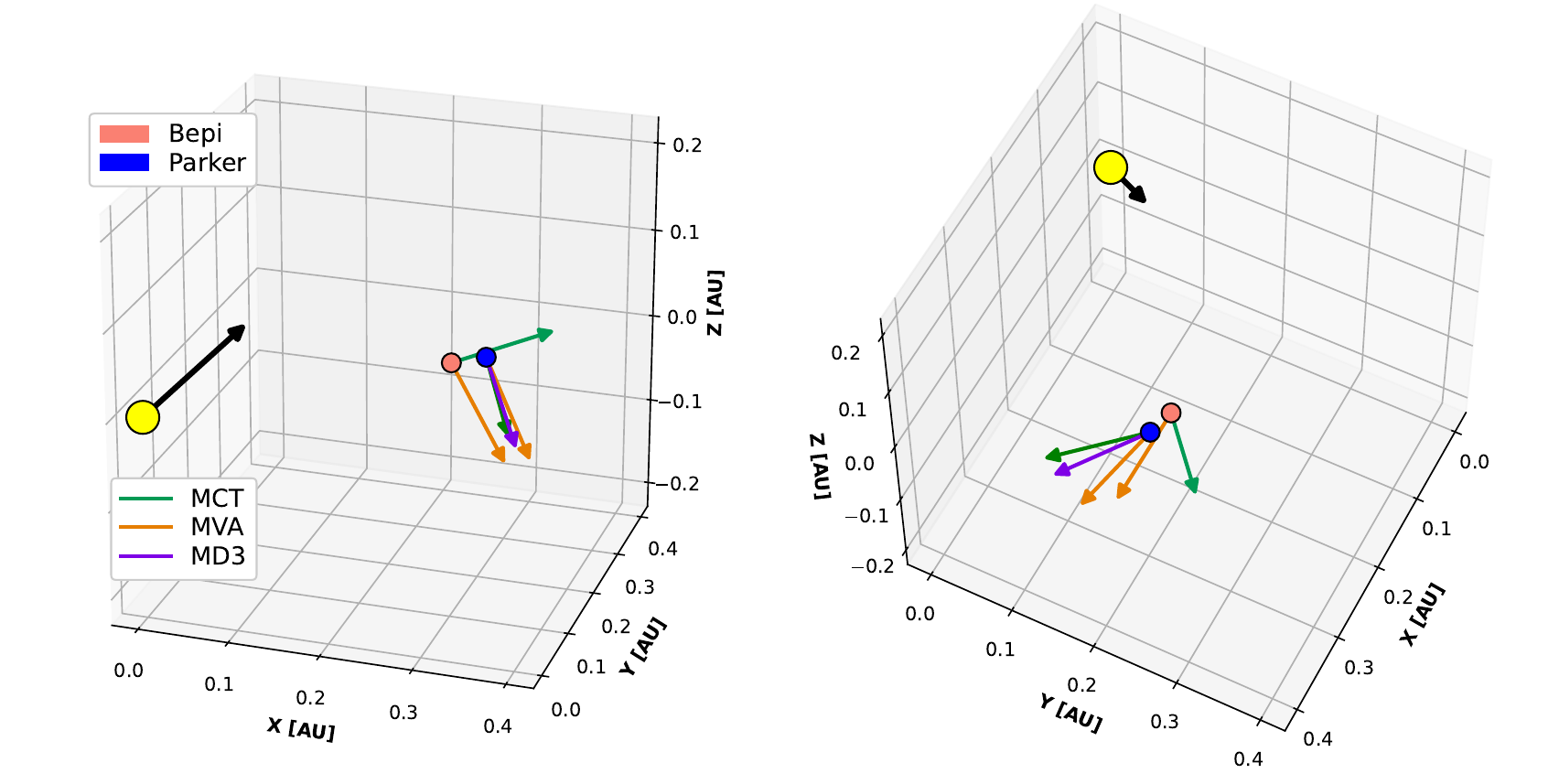}
\caption{Overview of the shock normals calculated with different methods at both Bepi and Parker, shown from two different perspectives to provide a better insight into their orientations in 3D. Normals calculated with the MCP and MVA methods are displayed for both spacecraft, whilst the normal derived using the MD3 technique is presented for Parker only since its derivation requires plasma data. The location of the Sun is represented with a yellow sphere, and the black arrow indicates the trajectory of the CME nose as derived from the GCS fitting shown in Figure~\ref{fig:wl_obs}.}
\label{fig:shocknormals}
\end{figure*}

The following parameters reported in Table~\ref{tab:shock} all concern the shock normal, $\mathbf{\hat{n}}$, and its related angles. Given the large uncertainties that are known to be associated with shock normal determination methods \citep[e.g.,][]{russell1983, schwartz1998}, we compare in this work results using three techniques: The magnetic coplanarity theorem \citep[MCT;][]{colburn1966}, minimum variance analysis \citep[MVA;][]{sonnerup1967}, and the mixed-mode method \citep[MD3;][]{abrahamshrauner1976}. We note that, whilst MCT and MVA are based uniquely on magnetic field data and can hence be applied to both probes, MD3 needs speed measurements and can thus be used at Parker only. The shock normals resulting from the different methods are visualised in 3D in Figure~\ref{fig:shocknormals}.

The MCT technique is based on the assumption that $\mathbf{B}_{u}$, $\mathbf{B}_{d}$, and $\mathbf{\hat{n}}$ lie in a single plane, and the shock normal is computed as

\begin{equation}
    \mathbf{\hat{n}}_\mathrm{MCT} = \pm  
    \frac{ (\mathbf{B}_{d} - \mathbf{B}_{u}) \times (\mathbf{B}_{d} \times \mathbf{B}_{u})}
    {|(\mathbf{B}_{d} - \mathbf{B}_{u}) \times (\mathbf{B}_{d} \times \mathbf{B}_{u})|} \, .
\label{eq:n_mct}
\end{equation}

The MVA technique assumes that, when a spacecraft crosses a ``transition layer'' (e.g., a shock front), the divergence-free nature of the magnetic field imposes the condition that the field component normal to such layer is continuous. It solves the set of three equations for the eigenvectors and eigenvalues of the magnetic variance matrix

\begin{equation}
    \sum_{\nu=1}^{3} M_{\mu\nu} n_{\nu} = \lambda n_{\mu} \, ,
\label{eq:n_mva}
\end{equation}

\noindent where $M_{\mu\nu} = \langle B_{\mu}B_{\nu}\rangle - \langle B_{\mu}\rangle \langle B_{\nu}\rangle$. The eigenvector associated with the smallest eigenvalue, $\lambda_{3}$, is the direction of minimum variance of the magnetic field and thus corresponds to the shock normal, $\mathbf{\hat{n}}_\mathrm{MVA}$. The reliability of the solution can be investigated via the intermediate-to-minimum eigenvalue ratio, $\lambda_{2}$/$\lambda_{3}$, and the minimum-variance-to-total magnetic field ratio across the discontinuity, $B_{n}$/$B$. Values of $\lambda_{2}$/$\lambda_{3} \geq$ 2 and $B_{n}$/$B \leq$ 0.3 are usually considered to describe the shock normal as well-defined \citep{lepping1980}.

Finally, the MD3 technique expands upon the MCT one by incorporating velocity changes across the shock. This method is usually considered more robust than those that only use magnetic field data, and the normal is calculated as

\begin{equation}
    \mathbf{\hat{n}}_\mathrm{MD3} = \pm  
    \frac{ (\mathbf{B}_{d} - \mathbf{B}_{u}) \times ((\mathbf{B}_{d} - \mathbf{B}_{u}) \times (\mathbf{V}_{d} - \mathbf{V}_{u}))}
    {|(\mathbf{B}_{d} - \mathbf{B}_{u}) \times ((\mathbf{B}_{d} - \mathbf{B}_{u}) \times (\mathbf{V}_{d} - \mathbf{V}_{u}))|} \, .
\label{eq:n_md3}
\end{equation}

Once the shock normals have been retrieved for each method and for each spacecraft, we calculate three angles related to them. The first is the shock angle, $\Theta$, between the shock normal and the upstream magnetic field, defined as 

\begin{equation}
    \Theta = \arccos \left( \frac{|\mathbf{B}_u\cdot\mathbf{\hat{n}}|}
    {\lVert\mathbf{B}_u\rVert \lVert\mathbf{\hat{n}}\rVert} \right) \, .
\label{eq:theta}
\end{equation}

The second angle is the so-called location angle, $\Lambda$, between the shock normal and the radial direction \citep{janvier2014}. It provides information as to the spacecraft crossing location with respect to the shock nose. Ideally, $\Lambda$ = 0$^{\circ}$ for a central encounter, and the larger the value (up to $|\Lambda|$ = 90$^{\circ}$), the farthest the crossing location is from the apex. The angle can be computed as

\begin{equation}
    \Lambda = \arctan \left( \frac{\sqrt{n_T^2 + n_N^2}}{n_R} \right) \, .
\label{eq:lambda}
\end{equation}

The third and final angle is the elevation angle, $\Xi$, between the shock normal and the RT plane (in RTN coordinates), defined as

\begin{equation}
    \Xi = \arctan \left( \frac{n_N}{\sqrt{n_R^2 + n_T^2}} \right) = \arcsin \left( \frac{n_{N}}{\lVert\mathbf{\hat{n}}\rVert} \right) \, .
\label{eq:elevation}
\end{equation}

According to the shock normal results shown in Table~\ref{tab:shock} and Figure~\ref{fig:shocknormals}, all the calculated normals point in a southeasterly way ($n_{y} <$ 0 and $n_{z} <$ 0) apart from the MCT case for Bepi, which points towards the northwest ($n_{y} >$ 0 and $n_{z} >$ 0). The shock is moderately quasi-parallel at Parker (31$^{\circ} < \Theta <$ 43$^{\circ}$), whilst for Bepi the MCT method retrieves a strongly quasi-parallel shock ($\Theta$ = 7.3$^{\circ}$) and the MVA technique produces a strongly quasi-perpendicular shock ($\Theta$ = 80.4$^{\circ}$). We note that quasi-parallel shocks are expected to occur more often closer to the Sun ($<$1~au) and quasi-perpendicular ones are more common at larger distances (${\sim}$1~au) due to the configuration of the Parker spiral \citep[e.g.,][]{richter1985, good2020}. Regardless of the normal direction and nature of the shock, all the obtained location angles suggest spacecraft crossings considerably far from the shock nose (40$^{\circ} < \Lambda <$ 65$^{\circ}$), as expected from our remote-sensing analysis of the 2022 February 15 CME (see Section~\ref{sec:erupt}).

Whilst the normals retrieved at Parker are all in general agreement, we consider the ones retrieved with the MCT and MD3 methods to be more reliable, given the large value of the $B_{n}$/$B$ ratio obtained with the MVA technique. At Bepi, on the other hand, both MVA ratios suggest a robust solution, which is in turn profoundly different from the one retrieved with the MCT method. Given the idealised (semi-spherical) shock geometry expected from the CME propagating to the northwest of the two probes (see Figure~\ref{fig:shocknormals}), it follows that the normal should indeed point approximately towards the southeast at the location of the spacecraft and, thus, one may intuitively favour the MVA results. Additionally, the uncertainties associated with the MCT method are known to become larger the closer a shock is to being fully parallel or fully perpendicular \citep{vinas1986}. However, given the prominent discrepancy between the two methods, we conclude that it is not possible to estimate a shock normal at Bepi in a reliable way. Nevertheless, it is clear that, despite the spatial proximity of Bepi and Parker, the two sets of measurements of the shock passage are dissimilar enough to produce fundamentally different results, possibly due to the shock being crossed significantly far from its nose and/or to an irregular shock front \citep[e.g.,][]{koval2010, kajdic2019} as well as the pre-existing local conditions of the turbulent solar wind \citep[e.g.,][]{guo2021, lavraud2021}.

\subsection{Sheath Region} \label{subsec:sheath}

The interplanetary shock detection was followed by a turbulent sheath region at both probes (see Figure~\ref{fig:insitu_mag}). We identify the sheath passage on 2022 February 16 during 06:25--15:41~UT at Bepi and 07:25--15:18~UT at Parker, resulting in a duration of 9.3~hr at the former and 7.9~hr at the latter. We note that both these values are significantly larger than what is expected at Mercury's orbit, where sheath passages have been found to last on average 2.4~hr by \citet{janvier2019}, or 1.7~hr (2.2~hr) for fast (slow) CME drivers by \citet{salman2020a}. A possible explanation for the observed long duration is that the CME was encountered by both probes close to its flank (as also confirmed by the interplanetary shock analysis in Section~\ref{subsec:shock}), since the thickness of a sheath is supposed to increase from the nose of the driving ejecta towards its sides \citep[e.g.,][]{gopalswamy2009, kilpua2017}. In such a case, Bepi would have encountered the CME farther from its nose than Parker. The mean (maximum) magnetic field magnitude measured in the sheath is 43.7~nT (69.8~nT) at Bepi and 38.8~nT (54.4~nT) at Parker, with values at both spacecraft being noticeably lower than the 57.8~nT (84.6~nT) determined by \citet{winslow2015} via a statistical analysis of 61 CMEs detected at Mercury's orbit. Again, a possible reason for the lower-than-average field strength is the probes' trajectory through the CME away from its nose.

\begin{figure*}[t!]
\centering
\includegraphics[width=\linewidth]{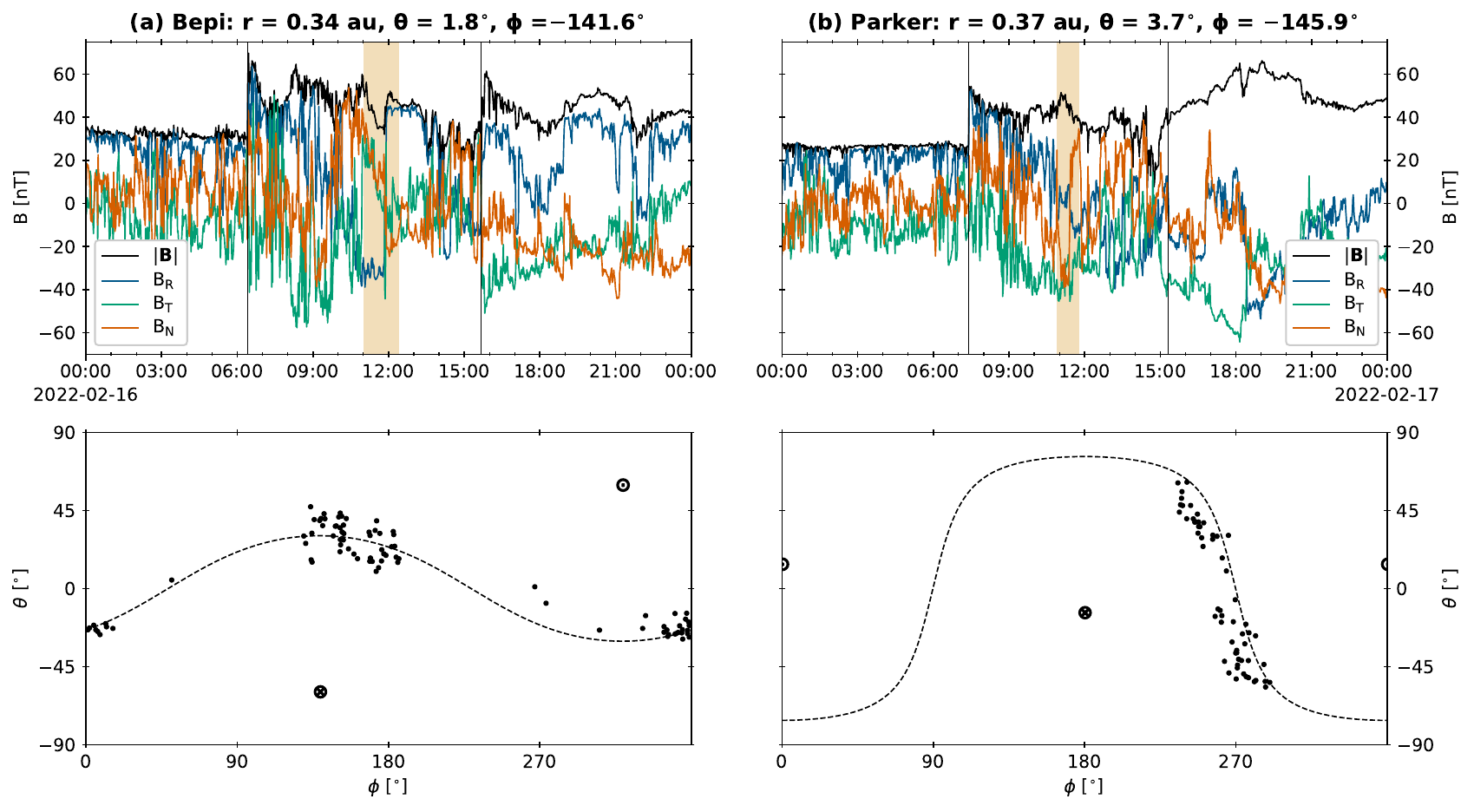}
\caption{The PMSs identified in the sheath regions at (a) Bepi and (b) Parker. The top panels show magnetic field measurements around the time of the sheath region passage (marked by the two vertical lines), with the PMS interval highlighted in okra. The bottom panels show the corresponding magnetic field vectors (within the PMS interval) on the $\theta$--$\phi$ surface, with the dashed curve representing the PMS plane and the $\odot$ ($\otimes$) symbol indicating the normal direction out of (into) it.}
\label{fig:pms_figure}
\end{figure*}

As mentioned in Section~\ref{subsec:mag}, the sheath region is the part displaying overall the highest magnetic field fluctuations in the time series shown in Figure~\ref{fig:insitu_mag} for either probe. In contrast to superposed epoch analyses performed on sheaths detected at 1~au, which display a clear (absolute) maximum at the shock passage and a local maximum at the ejecta leading edge time \citep[e.g.,][]{masiasmeza2016, kilpua2019b, salman2021}, the fluctuations observed at Bepi and Parker appear overall more uniformly distributed throughout the region. This may be due to the tendency for sheaths encountered closer to the Sun to generally display sharper field discontinuities and rotations \citep[see, e.g.,][who investigated the radial evolution of magnetic field fluctuation in sheath regions]{good2020}, or a result of passing through the flank of the CME. We remark, however, that statistical studies at 1~au often calculate $B_\mathrm{rms}$ over 1-min time scales, whilst here we considered 15-min intervals due to the available resolution of Bepi data for the period under analysis (see also Appendix~\ref{app:mpomag}). We find the average $B_\mathrm{rms}$ over the whole sheath region to amount to 5.3~nT at Bepi and 4.5~nT at Parker. Whilst this suggests a more turbulent sheath encountered at Bepi, it is important to bear in mind that this probe was closer to the Sun by ${\sim}0.03$~au in radial distance. If we consider, instead, the normalised $B_\mathrm{rms}$---obtained by dividing Equation~\ref{eq:rms} by the time-dependent magnetic field magnitude ($B$)---the average sheath fluctuations become ${\sim}0.13$ at either spacecraft, indicating similar conditions in both sets of measurements.

To further analyse the fine structure of the sheath regions observed at the two spacecraft, we investigate the presence (or lack thereof) of planar magnetic structures (PMSs) using the algorithm of \citet{palmerio2016}. PMSs are structures in the solar wind characterised by abruptly-changing magnetic field vectors that remain nearly parallel to a single plane \citep[e.g.,][]{nakagawa1989, neugebauer1993} and are frequently found in CME-driven sheath regions \citep[e.g.,][]{palmerio2016, ruan2023}. In brief, the PMS search algorithm of \citet{palmerio2016} first applies the MVA technique to the full sheath, verifying whether the PMS conditions of $\lambda_{2}$/$\lambda_{3} \geq 5$ \citep[e.g.,][]{savani2011} and $B_{n}$/$B \leq 0.2$ \citep[e.g.,][]{jones2000} are fulfilled, in which case the entire region is considered planar. If one or both criteria are not matched, then the algorithm scans through the sheath region using a ``sliding windows'' procedure. At every iteration, the sheath duration is reduced by 5~min from its end part, and the shortened window is shifted back towards the leading edge with 5-min increments, searching for the largest non-overlapping intervals for which the PMS criteria hold (note that a single sheath may contain more than one PMS). In the work of \citet{palmerio2016}, which analysed 95 sheaths observed near Earth, the minimum possible PMS duration was set to 1~hr. Here, to account for the fact that the structures under investigation were encountered at ${\sim}0.35$~au, where sheaths are on average more modest in size, we allow for a minimum PMS duration of 30~min. 

\begin{table}[t!]
\caption{Results of the PMS analysis at Bepi and Parker. \label{tab:pms} \vspace*{-0.12in}}
\centering
\renewcommand{\arraystretch}{1.05}
\begin{tabular}{l@{\hskip .32in}c@{\hskip .22in}c}
\toprule
 & \textsc{\textbf{BepiColombo}} & \textsc{\textbf{Parker Solar Probe}} \\
\midrule
$D$ [hr] & 1.4 & 1.0 \\
$\lambda_{2}$/$\lambda_{3}$ & 5.0 & 5.1 \\
$B_{n}$/$B$ & 0.10 & 0.12 \\
$O_\mathrm{PMS}$ & ($60^{\circ}$, $319^{\circ}$) & ($14^{\circ}$, $360^{\circ}$) \\
$\hat{n}_\mathrm{PMS}$ & [0.38, $-$0.33, 0.86] & [0.97, $-$0.00, 0.24] \\
\bottomrule
\end{tabular}
\vspace*{.1in}
\begin{tablenotes}
\item \emph{Notes.} $D$ = PMS duration. $\lambda_{2}$/$\lambda_{3}$ = eigenvalue ratio. $B_{n}$/$B$ = magnetic field ratio. $O_\mathrm{PMS}$ = orientation of the PMS plane in ($\theta$, $\phi$) format. $\hat{n}_\mathrm{PMS}$ = normal vector to the PMS plane.
\end{tablenotes}
\end{table}

The results of the PMS analysis are shown in Figure~\ref{fig:pms_figure} and Table~\ref{tab:pms}. The PMS search algorithm finds one planar structure lasting approximately one hour at the centre of each sheath. Despite their similar duration and location within their respective sheath regions, the two PMSs display different plane orientations, with the one found at Bepi oriented towards the northeast and the one found at Parker nearly aligned with the Sun--spacecraft line. Additionally, we note that the presence of a PMS in the central region of a sheath is a rather uncommon occurrence \citep[e.g.,][]{palmerio2016}, since planar structures are believed to form due to compression and alignment of discontinuities downstream of a shock \citep[e.g.,][]{jones2002} and/or due to magnetic field draping ahead of an ejecta \citep[e.g.,][]{kataoka2005}. Nevertheless, the excellent agreement between their respective positions within the sheath as well as their temporal duration suggests that the structures found at Bepi and Parker were the result of the same formation mechanism. Given the turbulent nature of the solar wind, especially in the downstream region of a shock, it is possible that the retrieved PMS orientations are deeply reflective of the local sheath conditions, and hence can vary more or less dramatically over relatively small spatial scales.

\begin{figure}[t!]
\centering
\includegraphics[width=\linewidth]{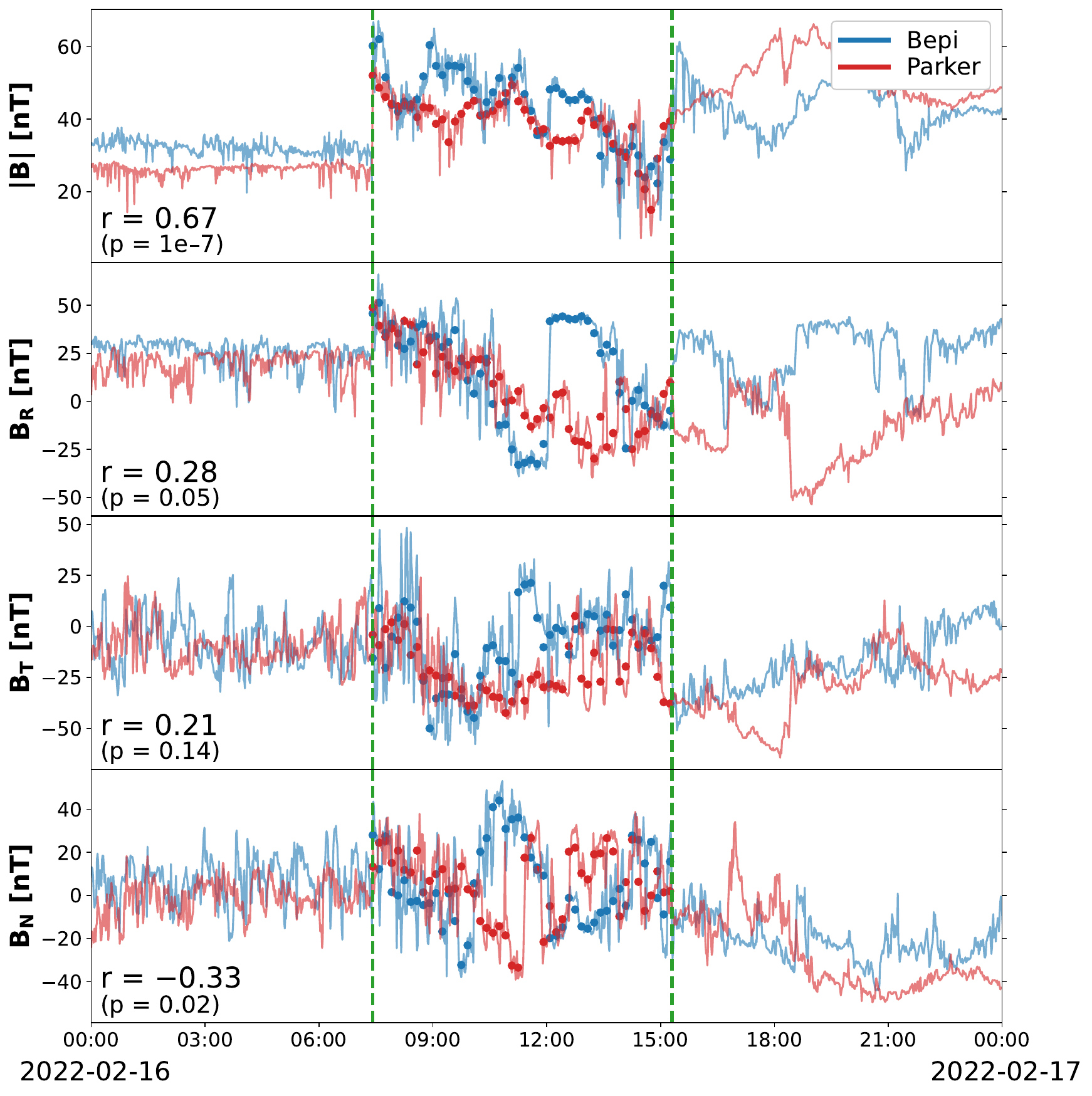}
\caption{Magnetic field observations of the sheath region driven by the 2022 February 15 CME at Bepi and Parker. The Bepi measurements have been stretched to match the sheath duration at Parker and time-shifted to align the two shock arrival times. The dashed vertical green lines mark the beginning and the end of the sheath region within which the magnetic field correlations are computed. The solid profiles show 1-min measurements, whilst the scatter points indicate the 10-min averaged data used to calculate the corresponding Pearson correlation coefficients (reported on the bottom left of each panel together with their p-values).}
\label{fig:correl_sheath}
\end{figure}

Indeed, this last point connects to the next step in our analysis, which is aimed at investigating the level of coherence of the sheath magnetic fields measured at the two spacecraft. To do so, we employ an approach similar to that used by \citet{alalahti2020}, who analysed the longitudinal spatial coherence of CME-driven sheath regions observed at 1~au by spacecraft characterised by non‐radial separations between 0.001 and 0.012~au ($0.06^{\circ}$--$0.69^{\circ}$). Specifically, they computed the Pearson correlation coefficients ($r$) between each pair of magnetic field components as well as the field magnitude. In the case explored here, the spacecraft non-radial separation is ${\sim}0.03$~au, which corresponds to ${\sim}5^{\circ}$ at ${\sim}0.35$~au. We note that, whilst this relative distance is larger than the range explored by \citet{alalahti2020}, it is well within the estimated scale lengths of typical magnetic field coherence in sheath regions resulting from their study (see also the discussion in the Introduction).

The results of the sheath magnetic coherence analysis are shown in Figure~\ref{fig:correl_sheath}. To directly compare the two sheath regions, we have stretched (or, actually, ``shrunk'') the Bepi measurements onto the Parker ones by aligning both the shock and the ejecta leading edge passage times. This approach is slightly different from the one employed by \citet{alalahti2020}, who examined the two magnetic field data sets using two methods: The first consists of maximising the cross-correlation of a combined Pearson coefficient that takes into account the magnetic field magnitude as well as its components, and the second consists of simply aligning the two shock arrival times (without stretching one time series onto the other). We have computed our correlations using the shock-aligning approach as well, and the results are qualitatively comparable to the ones displayed in Figure~\ref{fig:correl_sheath}. The Pearson correlation coefficients emerging from our analysis are [$B$, $B_{R}$, $B_{T}$, $B_{N}$] = [0.67, 0.28, 0.21, $-0.33$], hence, there is a good correlation in the magnetic field magnitude, but weak-to-no correlation in the field components. These results are in contrast to the ones from \citet{alalahti2020}, who reported the best correlation in the east--west component of the magnetic field ($B_{T}$) for sheath regions, with some degree of coherence maintained until ${\sim}0.15$~au (or ${\sim}9^{\circ}$) at 1~au. In fact, they suggested that correlations in total magnetic field ($B$) as well as its radial ($B_{R}$) and north--south ($B_{N}$) components are expected to cease for separations larger than 0.04~au (or ${\sim}2^{\circ}$) at 1~au. The higher-than-expected degree of coherence that we found in $B$ and $B_{R}$ for a ${\sim}5^{\circ}$ separation at ${\sim}0.35$~au might be related to the fact that changes in the sheath properties are known to increase with heliocentric distance as a CME propagates in interplanetary space \citep[e.g.,][]{salman2020a, winslow2021}. On the other hand, the lower-than-expected degree of coherence that we found in $B_{T}$ may be due to the different Parker spiral orientation at ${\sim}0.35$~au when compared to 1~au---i.e., characterised by a significantly smaller component in the east--west direction.

Finally, we inspect the available plasma data from Parker (see Figure~\ref{fig:insitu_psp}) to obtain a deeper insight into the sheath's structure and evolution, including the relationship with its driver. First of all, we focus on the speed profile measured within the sheath region and apply a linear regression to it, finding an overall decrease of ${\sim}3$\% from the interplanetary shock to the ejecta leading edge (speed decline by 26~km$\cdot$s$^{-1}$ and average speed of 901~km$\cdot$s$^{-1}$), with a slope relative to the average sheath speed of 0.37\%~hr$^{-1}$. The mean error associated with the regression, however, is 4.8\% with respect to the average sheath speed, making the structure encountered at Parker akin to the ``Category-D'' sheaths of \citet{salman2021}, i.e.\ those characterised by a complex (non-linear) speed profile. The statistical analysis of \citet{salman2021} was centred on sheaths preceded by shocks encountered at 1~au and, amongst their findings, the authors reported that Category-D sheaths tend to be driven by CMEs that are detected in situ far from their apex, as is indeed the case for the 2022 February 15 event.

Another interesting aspect to investigate is the contribution of CME propagation and CME expansion to the sheath formation---CME-driven sheaths, in fact, are considered ``hybrid'' structures that feature aspects of propagation sheaths, in which the solar wind flows around a propagating object, and expansion sheaths, in which the solar wind piles up ahead of an expanding object \citep[e.g.,][]{siscoe2008}. To do so, we calculate two Mach numbers: $M_\mathrm{prop}$, based on the ejecta propagation speed, and $M_\mathrm{exp}$, based on the ejecta expansion speed \citep[see Equations~1 and 2 in][for a definition of the two parameters]{salman2021}. The propagation speed is simply the ejecta average speed, $V_\mathrm{prop} = 589$~km$\cdot$s$^{-1}$, whilst the expansion speed is defined as the half-difference between the speeds of the leading and trailing edges \citep{owens2005}, $V_\mathrm{exp} = 142$~km$\cdot$s$^{-1}$. We obtain $M_\mathrm{prop} = 0.27$ and $M_\mathrm{exp} = 0.82$, indicating that the sheath driven by the 2022 February 15 CME was mainly an expansion one. This is consistent with the presence of a PUC region immediately ahead of the ejecta leading edge (see the prominent density enhancement in the trailing portion of the sheath in Figure~\ref{fig:insitu_psp}), usually attributed to strong CME overexpansion. In fact, the CME analysed in this work might still have been experiencing significant expansion by the time it was detected at Parker since the sheath appears overall less dense than the following ejecta, in agreement with \citet{salman2021} who found that expansion sheaths tend to display lower densities on average. Furthermore, \citet{temmer2022} estimated that the sheath density tends to overcome the ejecta density in the heliocentric distance range of 0.09--0.28~au, after which expansion of the driver generally weakens. \edit1{We also note that \citet{giacalone2023} came to the same conclusion (i.e., that the 2022 February 15 CME was overexpanding by the time it impacted Parker) by observing the intensity of energetic particles increase behind the shock, possibly resulting from ions filling an expanding volume associated with the propagation of a blast wave.}

\subsection{Magnetic Ejecta} \label{subsec:ejecta}

The passage of the sheath region was followed by the arrival of the magnetic ejecta at both probes (see Figure~\ref{fig:insitu_mag}). We determine the ejecta boundaries on 2022 February 16--17 during 15:41--07:51~UT at Bepi and 15:18--16:48~UT at Parker---noting that flux rope signatures at the latter spacecraft were identified until 06:32~UT on February 17 (see Section~\ref{sec:insitu}). These intervals result in ejecta durations of 16.2~hr at Bepi and 25.5~hr at Parker, i.e.\ significantly larger than the average of 7.2~hr reported by \citet{janvier2019} for CMEs encountered at ${\sim}$0.4~au. The mean (maximum) magnetic field magnitude measured in the ejecta is 41.3~nT (61.4~nT) at Bepi and 45.9~nT (66.3~nT) at Parker, noting that values at both spacecraft are lower than the average values at Mercury's orbit of 55.3~nT (86.2~nT) as reported by \citet{winslow2015}. \edit1{Additionally, we note that both the mean and maximum field values at Parker are higher than at Bepi, despite the former being slightly farther from the Sun than the latter. This may be due to the longitudinal and latitudinal separation between the two probes (i.e., to intrinsic differences in the CME structure over relatively short spatial scales) and/or to the higher-speed wind following the ejecta at Parker, resulting in compression of the field.} Overall, possible reasons for the longer durations and lower field magnitudes measured in the two ejecta investigated here \edit1{in comparison to average values at Mercury} are the fact that the event was encountered in situ only through its southern flank (see Section~\ref{sec:erupt}) as well as the strong expansion experienced by the CME---as mentioned in Section~\ref{subsec:sheath}, we find an expansion speed (based on Parker measurements), $V_\mathrm{exp}$, of ${\sim}142$~km$\cdot$s$^{-1}$. When considering the magnetic cloud interval only, the expansion speed is basically unchanged (${\sim}144$~km$\cdot$s$^{-1}$), since the trailing portion of the ejecta features a plateau in the bulk velocity. These values are significantly higher than the average CME expansion speeds of ${\sim}30$~km$\cdot$s$^{-1}$ found at 1~au \citep[e.g.,][]{nieveschinchilla2018a, lugaz2020}, but it remains unclear what the corresponding ``typical'' expansion speeds at ${\sim}$0.35~au may be due to the lack of systematic in-situ plasma measurements---although it is generally assumed that CME expansion should somewhat decrease between the Sun and Earth's orbit \citep[e.g.,][]{wang2005, zhuang2023}.

\begin{figure*}[th!]
\centering
\includegraphics[width=\linewidth]{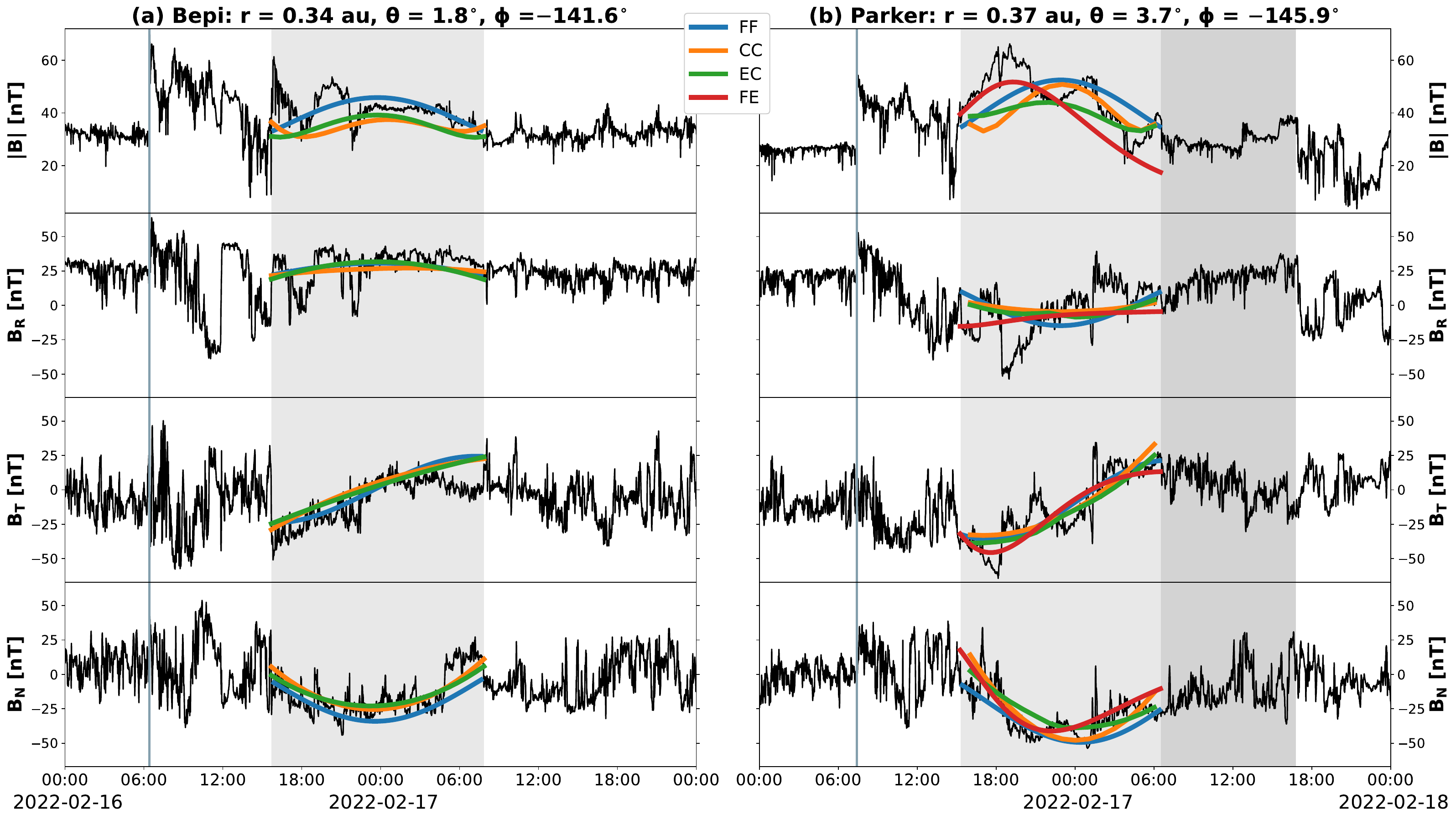}
\caption{Flux rope fitting results at (a) Bepi and (b) Parker, showing magnetic field data at each spacecraft and the corresponding quantities resulting from different flux rope models. Note that, at Parker, the fitting is applied to the magnetic cloud interval only (see Appendix~\ref{app:psp_fits} for the corresponding results when considering the full magnetic obstacle).}
\label{fig:fr_fits}
\end{figure*}

\begin{table*}[t!]
\caption{Flux rope fitting results at Bepi and Parker. \label{tab:fr_fits} \vspace*{.1in}}
\centering
\renewcommand{\arraystretch}{1.05}
\hspace*{-.6in}
\begin{tabular}{l@{\hskip .5in}ccccccc}
\toprule
& \multicolumn{2}{c}{\textbf{FF}} & \multicolumn{2}{c}{\textbf{CC}} & \multicolumn{2}{c}{\textbf{EC}} & {\textbf{FE}}\\
& \textbf{Bepi} & \textbf{Parker} & \textbf{Bepi} & \textbf{Parker} & \textbf{Bepi} & \textbf{Parker} & \textbf{Parker} \\
\midrule
$H$ & +1 & +1 & +1 & +1 & +1 & +1 & +1 \\
$\vartheta_{0}$ [$^{\circ}$] & $-78.8$ & $-29.0$ & $-72.5$ & $-66.8$ & $-63.4$ & $-51.1$ & $-53.2$\\
$\varphi_{0}$ [$^{\circ}$] & 167.5 & 190.8 & 64.0 & 251.7 & 15.0 & 233.1 & 290.7 \\
$B_{0}$ [nT] & 63.8 & 66.5 & 60.1 & 51.0 & 50.3 & 43.8 & 76.2 \\
$R_{0}$ [au] & 0.166 & 0.074 & 0.165 & 0.108 & 0.145 & 0.115 & 0.101\\
$p_{0}$ [$R_{0}^{-1}$] & 0.681 & 0.573 & 0.715 & 0.052 & 0.588 & 0.178 & 0.396 \\
$\tau$ [au$^{-1}$] & 5.21 & 11.69 & 0.53 & 0.95 & 0.71 & 2.05 & 8.54 \\
$\Phi_\mathrm{ax}$ [10$^{21}$ Mx] & 5.35 & 1.10 & 6.31 & 2.31 & 3.91 & 1.21 & 2.39 \\
$\Phi_\mathrm{po}$ [10$^{21}$ Mx/au] & 9.89 & 4.57 & 5.94 & 3.97 & 4.87 & 3.76 & 7.20 \\
C1 & --- & --- & 1.71 & 1.42 & 1.50 & 0.96 & --- \\
$\psi$ [$^{\circ}$] & --- & --- & --- & --- & 19.4 & 74.3 & --- \\
$\delta$ & --- & --- & --- & --- & 0.96 & 0.54 & --- \\
\bottomrule
\end{tabular}
\vspace*{.1in}
\begin{tablenotes}
\item \emph{Notes.} $H$: helicity sign (or chirality); $\vartheta_{0}$: axis latitude; $\varphi_{0}$: axis longitude; $B_{0}$: axial magnetic field magnitude; $R_{0}$: flux rope radius; $p_{0}$: normalised impact parameter; $\tau$: twist; $\Phi_\mathrm{ax}$: axial flux; $\Phi_\mathrm{po}$: poloidal flux; C1: ratio of the azimuthal-to-axial current at the flux rope outer boundary; $\psi$: propagation angle (i.e., rotation around the flux rope axis); $\delta$: cross-sectional distortion (equals 1 for a circular cross-section and 0 for maximum distortion). The assumed average speed for the CME ejecta in \edit1{the FF, CC, and EC fits} is 654~km$\cdot$s$^{-1}$, \edit1{whilst the FE procedure uses a polynomial fit to the observed speed profile}.
\end{tablenotes}
\end{table*}

To obtain an indication of the overall cross-sectional ejecta structure encountered by either probe, we evaluate the distortion parameter (DiP) introduced by \citet{nieveschinchilla2018a}. In brief, the DiP quantifies the level of distortion in a CME, due to e.g.\ interactions with the background solar wind, by examining the measured profile of the magnetic field magnitude, and is defined as the fraction of the full ejecta interval where 50\% of the total $B$ is accumulated. We find a DiP at Bepi of 0.48, whilst at Parker we obtain DiP = 0.39 for the magnetic obstacle and DiP = 0.43 for the magnetic cloud interval. According to \citet{nieveschinchilla2018a}, events with DiP = 0.50 $\pm$ 0.07 can be considered symmetric in their magnetic field profile, which is the case for Bepi according to our results. For Parker, the magnetic obstacle interval features a DiP that is consistent with compression at the front and that, together with the high $V_\mathrm{exp}$ mentioned above, indicates that the asymmetric nature of the $B$ profile is due to expansion of the structure as it crosses the spacecraft---also known as the `ageing effect' \citep[e.g.,][]{demoulin2008}. However, if we consider only the magnetic cloud portion of the ejecta at Parker, we find a DiP value that suggest an approximately-symmetric profile alongside a high $V_\mathrm{exp}$. At the same time, it is unlikely that the large speed gradient (between the ejecta leading and trailing edges) found at Parker corresponds to a flat speed profile at Bepi. \citet{nieveschinchilla2018a}, in their study of 337 CMEs detected at 1~au, found only a few events characterised by 0.43 $\leq$ DiP $\leq$ 0.57 and $V_\mathrm{exp} \geq 100$~km$\cdot$s$^{-1}$, contradicting the expected effect of CME expansion on the magnetic field strength. It is unclear why this is the case here, but one possibility may be local distortions in the structure---see, e.g., how both sets of measurements in Figure~\ref{fig:insitu_mag} feature irregular profiles of the magnetic field magnitude, with several ``bumps'' across the ejecta passage, and how the plasma data at Parker in Figure~\ref{fig:insitu_psp} display complex trends in the density and temperature.

\begin{figure*}[th!]
\centering
\includegraphics[width=\linewidth]{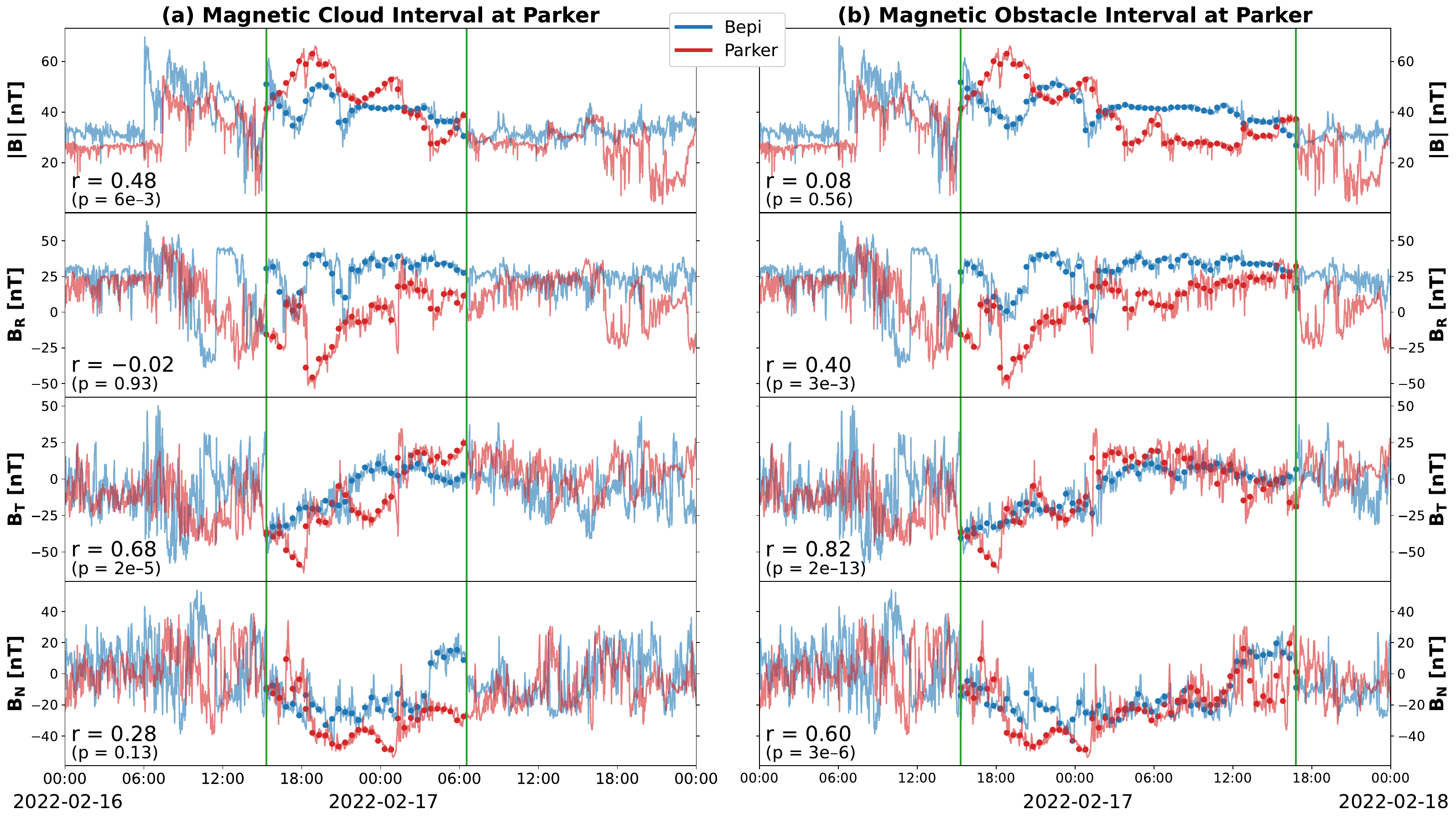}
\caption{Magnetic field observations of the 2022 February 15 CME ejecta at Bepi and Parker. The Bepi measurements have been stretched to match the durations of (a) the magnetic cloud and (b) the magnetic obstacle at Parker and time-shifted to align the two ejecta leading edge times. The dashed vertical green lines mark the leading and trailing edges within which the magnetic field correlations are computed. The solid profiles show 1-min measurements, whilst the scatter points indicate the 30-min averaged data used to calculate the corresponding Pearson correlation coefficients (reported on the bottom left of each panel together with their p-values).}
\label{fig:correl_ejecta}
\end{figure*}

To evaluate the large-scale internal magnetic structure of the ejecta encountered at Bepi and Parker we apply three different flux rope fitting models to the in-situ magnetic field measurements. Specifically, we employ the force-free constant-$\alpha$ \citep[FF;][]{lepping1990}, the circular-cylindrical \citep[CC;][]{nieveschinchilla2016}, and the elliptic-cylindrical \citep[EC;][]{nieveschinchilla2018b} analytical descriptions of flux ropes to recover geometric and magnetic parameters that allow us to estimate orientation, size, and flux content. All these methods require an average CME ejecta speed, for which we employ a value of 654~km$\cdot$s$^{-1}$ based on the speed profile of the magnetic cloud portion at Parker (under the assumption that Bepi encountered similar velocities). \edit1{Additionally, we employ the linear force-free self-similarly expanding cylindrical flux rope model \citep[FE;][]{marubashi2007}, which is, in other words, an FF model that includes expansion. The FE technique requires fitting a full speed profile alongside the magnetic field components and, thus, is applicable in this case only to Parker data---here, we obtain our speed profile via a second-order polynomial fit to the in-situ measurements.} Results applied to the (full) ejecta at Bepi and the magnetic cloud portion of the ejecta at Parker are shown and summarised in Figure~\ref{fig:fr_fits} and Table~\ref{tab:fr_fits}. The corresponding fits for the magnetic obstacle interval at Parker are reported and discussed in Appendix~\ref{app:psp_fits}. 

We note that, despite the different fits appearing overall consistent at either spacecraft from visual inspection of Figure~\ref{fig:fr_fits}, the results reported in Table~\ref{tab:fr_fits} display some stark differences, in terms of both geometric and magnetic parameters. Nevertheless, we can extrapolate some common trends: For example, all techniques recover a right-handed flux rope at both probes (as expressed by the positive helicity sign, $H$), determine a higher axis inclination at Bepi when compared to Parker (as expressed by the latitude angle, $\vartheta_{0}$), and estimate a more central encounter at Parker as opposed to Bepi (as expressed by the impact parameter, $p_{0}$). It has been shown that flux rope fitting results using different techniques can feature some disagreement even in ``simple'' cases characterised by negligible expansion speeds and symmetric magnetic field strength profiles \citep[e.g.,][]{alhaddad2018}, hence it is not surprising to find discrepancies in this complex, flank-encounter event. \edit1{One example can be seen in the $B_R$ profiles shown in Figure~\ref{fig:fr_fits}, which display, on average, opposite polarities at the two spacecraft. If a different flux rope model were used that fit the $B_R$ component with a more linear profile \citep[such as the uniform-twist geometry; e.g.,][]{farrugia1999}, there could perhaps be less disagreement between the orientations obtained for each spacecraft. Nevertheless, flank encounter trajectories tend to be especially challenging for all in-situ flux rope models \citep[cf.\ with the ``problematic cases'' discussed by][]{lynch2022}.} Overall, the most important conclusion to draw is that each of the flux rope models, when considered individually, returns very different ``best fit'' parameter sets for the large-scale structure observed at Bepi and Parker, reflecting the significant differences in the (local) magnetic field time series measured within the ejecta by each spacecraft. We remark that the same holds true when considering the magnetic obstacle interval at Parker, as seen in the results presented in Appendix~\ref{app:psp_fits}.

To further investigate the coherence of the magnetic fields measured by the two probes, we analyse the corresponding correlations in the observed time series by adopting a similar approach to \citet{lugaz2018}, who compared the ejecta of CMEs measured near 1~au over non-radial separations of 0.005--0.012~au ($0.29^{\circ}$--$0.69^{\circ}$). As was the case for the coherence analysis in the sheath, the non-radial separation between Bepi and Parker of ${\sim}0.03$~au (${\sim}5^{\circ}$) during 2022 February 16--17 is larger than the range explored by \citet{lugaz2018}, but falls well within the typical scales of magnetic field coherence emerging from their study (see also the discussion in the Introduction). The results of our analysis are shown in Figure~\ref{fig:correl_ejecta}. In their study, \citet{lugaz2018} aligned the magnetic field profiles by calculating the lag between the pairs of data that maximised the Pearson correlations, with the lags being computed separately for the field magnitude and each of the three components. Here, on the other hand, we adopt the time series stretching approach that we employed for the sheath region in Section~\ref{subsec:sheath}, allowing us to compare directly the magnetic fields measured at Bepi with those detected at Parker in the magnetic cloud (Figure~\ref{fig:correl_ejecta}(a)) as well as in the magnetic obstacle (Figure~\ref{fig:correl_ejecta}(b)) intervals. By doing so, we find an intriguing result: In fact, the Pearson correlation coefficients are [$B$, $B_{R}$, $B_{T}$, $B_{N}$] = [0.48, $-0.02$, 0.68, 0.28] when considering the Parker magnetic cloud only, but change dramatically into [$B$, $B_{R}$, $B_{T}$, $B_{N}$] = [0.08, 0.40, 0.82, 0.60] when computed over the Parker magnetic obstacle interval. Hence, for the magnetic cloud we find good correlation in $B_{T}$, moderate correlation in $B$, and weak-to-no correlation in $B_{R}$ and \edit1{$B_{N}$}, whilst for the magnetic obstacle we obtain moderate-to-strong correlation in all the components with no correlation in the magnitude. 

\begin{figure*}[t!]
\centering
\includegraphics[width=\linewidth]{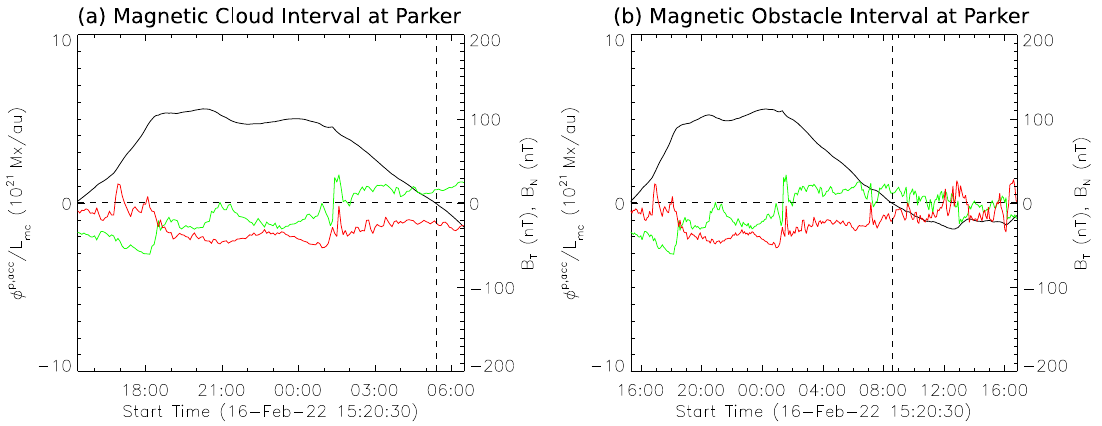}
\caption{Overview of the flux erosion analysis for the 2022 February 15 CME ejecta at Parker, showing time series of the accumulated poloidal flux (black) as well as the magnetic field components $B_{T}$ (green) and $B_{N}$ (red). The same analysis is performed for the (a) magnetic cloud and (b) magnetic obstacle intervals. The dashed vertical line marks the time at which flux asymmetry is estimated to begin.}
\label{fig:erosion}
\end{figure*}

These results are in contrast to the ones from \citet{lugaz2018}, who found the greatest correlations in the magnetic field strength ($B$) and estimated that some degree of coherence should be maintained up to separations of 0.25--0.35~au ($15^{\circ}$--$21^{\circ}$) at 1~au---for the single components, typical coherence scales lie instead between 0.07--0.12~au ($4^{\circ}$--$7^{\circ}$). According to our results, the highest degree of coherence is obtained in the east--west ($B_{T}$) component regardless of the interval considered at Parker, and in the case of the magnetic obstacle interval it is possible to encounter no correlation in the field magnitude whilst, at the same time, the single components all display moderate-to-strong agreement. A reasonable explanation for the generally lower-than-expected degree of coherence found in the 2022 February 15 event is that the CME was encountered far from its nose and with a high impact parameter by both spacecraft, thus considerably away from the stronger, possibly more ordered inner layers of the embedded flux rope. Furthermore, we note that \citet{lugaz2018} deemed the correlation coefficient for the radial component of the magnetic field ($B_{R}$) ill-defined due to the lack of variation that is often observed along the Sun--observer line in flux rope-like magnetic obstacles. In the event studied here, $B_{R}$ is comparable in intensity to the other components, hence its lower correlation coefficient when compared to $B_{T}$ and $B_{N}$ (regardless of the interval under consideration at Parker) appears to be due to intrinsic differences between the two ejecta---at Bepi, $B_{R}$ is predominantly positive, whilst the opposite holds true at Parker. Again, this may be due to the glancing nature of the spacecraft encounters under investigation.

The magnetic field correlation analysis reported above suggests that, indeed, the ``longer'' interval that we considered at Parker corresponds to the ``original'', full CME ejecta. Nevertheless, from all the instances in which these data are shown (Figures~\ref{fig:insitu_mag}, \ref{fig:insitu_psp}, \ref{fig:fr_fits}, and \ref{fig:correl_ejecta}) it is clear that there is a transition between the two distinct magnetic field environments of what we have defined as the magnetic cloud and the trailing portion of the magnetic obstacle. In Section~\ref{subsec:psp}, we tentatively attributed these features to magnetic flux erosion, which is a consequence of CME reconnection with the surrounding solar wind. Here, we investigate whether this is indeed the case. Figure~\ref{fig:erosion} provides an overview of the results emerging from our analysis aimed at quantifying the amount of poloidal flux eroded due to reconnection in the interplanetary medium (i.e., shortly before the CME was detected in situ). These results employ as input outputs from the EC model, which allows for elliptical, distorted cross-sections and thus has more degrees of freedom than the remaining two techniques showcased in Figure~\ref{fig:fr_fits}---we remark that we have repeated the same analysis using instead the FF model with expansion (FE) introduced in Appendix~\ref{app:psp_fits}, and results are very well consistent with the EC ones. Once geometric parameters (from any fitting model) such as flux rope axis direction, radius, and impact parameter are provided, we calculate the inbound--outbound flux asymmetry using the ``direct method'' of \citet{dasso2005} as described in \citet{pal2022}. We perform these computations for both the magnetic cloud (Figure~\ref{fig:erosion}(a)) and the full magnetic obstacle (Figure~\ref{fig:erosion}(b)) intervals. According to the results shown in Figure~\ref{fig:erosion}, flux asymmetry is found at the back of the ejecta in both cases, indicating reconnection and erosion at the front of the rope \citep[see the scenario depicted in Figure~6 of][]{pal2022}. The asymmetry starts on 2022 February 17 at 05:25~UT (08:35~UT), and the eroded poloidal flux is $1.4\times10^{21}$~Mx$\cdot$au$^{-1}$ ($1.5\times10^{21}$~Mx$\cdot$au$^{-1}$) for the shorter (longer) interval \citep[noting, however, that the glancing nature of the encounter may introduce additional errors in the results; e.g.,][]{ruffenach2015}. Our ``by-eye'' estimate of such boundary fell on February 17 at 06:32~UT, i.e.\ comfortably in between the two times obtained with the flux erosion computation method. Thus, what emerges from this investigation is that the ejecta encountered at Parker was eroded, which may explain (at least in part) some of the differences in the structures encountered at the two spacecraft.


\section{Discussion} \label{sec:discuss}

The analysis and results presented in Section~\ref{sec:analysis} reveal a series of similarities but also differences in the local measurements of the 2022 February 15 CME from Bepi and Parker, despite their relative proximity (${\sim}$0.03~au in radial distance and ${\sim}$5$^{\circ}$ in non-radial separation). The eruption responsible for the event observed in situ was an impressive filament eruption that resulted in a fast (${\sim}2200$~km$\cdot$s$^{-1}$) CME \citep[see][]{mierla2022}, observed in remote-sensing imagery to propagate with a significant northward component and thus expected to impact any in-situ, near-the-ecliptic location through its southern flank. Here, we shall summarise the main findings with respect to the main structures that have been investigated, namely the interplanetary shock, the sheath region, and the magnetic ejecta.

Analysis of the CME-driven interplanetary shock (Section~\ref{subsec:shock}) revealed that the two data sets featured an equal magnetic compression ratio, but determination of the shock normal and related parameters was inconsistent between the two probes. Whilst at Parker the three employed methods (MCT, MVA, and MD3) were generally in agreement in determining a moderately quasi-parallel shock with its normal pointing approximately towards the southeast, at Bepi the two adopted techniques (MCT and MVA, since plasma data necessary for MD3 were not available) provided two extremely different results: one suggesting a strongly parallel shock and the other indicating a strongly perpendicular one, with contrasting normal directions. It is interesting to note that the MVA method at the two spacecraft provided shock normal directions that are almost identical, but resulted in an intermediate shock at Parker and a perpendicular one at Bepi. Overall, we concluded that the shock properties at Parker were at least qualitatively well-defined, whilst at Bepi it was not possible to obtain reliable results. It is likely that availability of plasma data at Bepi would have aided towards the solution of the encountered discrepancies: apart from allowing determination of other properties such as density compression ratio and velocity jump, including speed changes across the discontinuity in shock normal estimations is known to provide more robust results. The shock normal directions determined using all methods at Parker and MVA at Bepi---i.e., roughly southeasterly---were at least consistent with the idealised scenario of a bubble-like shock surface that propagated to the northwest of the in-situ observers.

Analysis of the CME-driven sheath region (Section~\ref{subsec:sheath}) showed that the two spacecraft detected a comparable amount of magnetic field fluctuations on average, indicating similar conditions of the corresponding magnetic environment. Another encountered similarity was that both sheaths had a PMS embedded in its middle, although the resulting PMS plane displayed different orientations. We speculated that the PMSs at the two probes formed via the same mechanism, and that the obtained orientations reflected the turbulent nature of the local (pre-shock-passage) solar wind. A direct comparison of the magnetic field measurements within the two sheath region data sets showed good correlation in the field magnitude, but weak-to-no correlation in the field components. It is well known that CME-driven sheaths are highly fluctuating structures \citep[e.g.,][]{moissard2019, kilpua2020}, hence variations over relatively close locations may be expected. For example, \citet{kilpua2021} found, in a sheath measured at 1~au over distances of $0.01$~au in the radial direction, $0.5^{\circ}$ in latitude, and $1.4^{\circ}$ in longitude, different sheath durations, different amounts and positioning of magnetic field discontinuities within the sheath, and even different locations within the structure of an embedded small-scale flux rope. A more complete understanding of sheath region formation, evolution, and 3D variation will require multi-spacecraft in-depth analyses of CME-driven sheaths during periods featuring radially and/or longitudinally aligned probes \citep[e.g.,][]{good2020, salman2020a}.

Analysis of the CME magnetic ejecta (Section~\ref{subsec:ejecta}) showed a single, relatively coherent structure at Bepi, but a ``two-part ejecta'' at Parker---we identified the leading portion as the magnetic cloud (corresponding to the ``core'' flux rope), and the higher-fluctuating trailing part as reconnected (open) field signature of erosion at the front of the ejecta. We employed different flux rope fitting techniques to recover geometric and magnetic properties of the ejecta measured at Bepi and Parker---considering both the magnetic cloud and the full magnetic obstacle intervals at the latter---and found more or less dramatic difference not only between the different models, but also between the two spacecraft within a single model. It is well known that different flux rope fitting models applied to the same data set may display significant discrepancies, especially in the case of flank encounters with high impact parameters \citep[e.g.,][]{riley2004, alhaddad2013}. In the case investigated here, one may have expected better agreement due to the measurements being taken significantly closer to the Sun than 1~au, i.e.\ where less CME evolution has occurred. However, \citet{lynch2022} showed that flux rope fitting models perform generally worse for flank encounters even as close as 10--30\,$R_{\odot}$ using results from a magnetohydrodynamic model, which are characterised by smoother fields and do not fully reproduce the highly turbulent nature of the solar wind. Nevertheless, most fitting techniques (with the notable exception of the FF model at Parker) recovered a high-inclination flux rope, in agreement with GCS reconstruction results based on remote-sensing coronagraph observations (see Figure~\ref{fig:wl_obs}) and suggesting that the CME as a whole largely maintained its orientation as it travelled to ${\sim}0.35$~au. The magnetic field correlation analysis between the two ejecta (comparing Bepi with both ejecta ``options'' at Parker) showed the highest agreement for the $B_{T}$ component, whilst correlation in the field strength and the remaining two components varied depending on the ejecta interval under consideration at Parker. In particular, it was interesting to observe moderate-to-high correlation in all the components, but no correlation in the magnitude for the magnetic obstacle interval at Parker.

Overall, our investigation of the 2022 February 15 CME measured in situ at two relatively close spacecraft revealed some significant (local) variations of its large-scale structure. For example, the dramatically different duration of the ejecta interval---16.2~hr at Bepi and 25.5~hr at Parker, for a total difference of 9.3~hr---cannot be explained uniquely in light of CME expansion for a radial separation between the two probes of 0.03~au. \edit1{We note that \citet{regnault2023} reached the same conclusion for an event observed at SolO and Earth (separated at the time by 0.13~au in heliocentric distance and ${\sim}2^{\circ}$ in angular distance), highlighting how small spatial separations can have a prominent impact on the observed CME properties.} Likewise, the different flux rope orientations recovered via fitting models (keeping in mind, however, the large uncertainties involved, \edit1{especially at Bepi due to the extra data processing described in Appendix~\ref{app:mpomag}}) suggest, in agreement with previous work \citep[e.g.,][]{farrugia2011, mostl2012, pal2023}, that the axial field direction is more of a ``local'' quantity rather than part of a coherent, large-scale flux tube. In the case analysed here, it is possible that interaction with a following HPS had some influence on the orientation of the ejecta at Parker, as e.g.\ the events investigated by \citet{farrugia2011} and \citet{pal2023} involved interactions with stream interaction regions or those studied by \citet{mostl2012} involved CME--CME interactions. Finally, it is interesting to speculate on the uniqueness of the 2022 February 15 event that was the focus of this work. At both probes, we found average shock strengths, as well as longer durations and lower magnetic field magnitudes in both the sheath and ejecta when compared to typical values measured at Mercury's orbit. The strong expansion that we found based on Parker observations is in agreement with \citet{scolini2021}, who found that CMEs tend to expand rapidly until 0.4~au and shift to a more moderate growth rate at larger distances. Altogether, these features are consistent with those of a rather strong CME encountered towards its flank, where magnetic and kinematic properties are generally weaker. A follow-up investigation that could significantly complement the present work should compare these results with observations of a central, near-nose encounter under similar positioning conditions of the involved spacecraft. It is possible, in fact, that sampling two close-by trajectories closer to a flux rope central axis would result in much more consistent in-situ data sets, that would in turn inform us as to the expected coherence of CMEs across their full 3D structure. However, such a study would have to rely, as this one, on a fortuitous spacecraft conjunction together with a CME launched in a near-optimal direction, the likeliness of which can be estimated at the moment only in probabilistic terms.


\section{Summary and Conclusions} \label{sec:conclu}

In this work, we have analysed in detail for the first time the mesoscale structure of a CME that was detected at Mercury's orbit by two different probes, i.e.\ Bepi and Parker, that were separated by ${\sim}$0.03~au in radial distance, ${\sim}$2$^{\circ}$ in latitude, and ${\sim}$4$^{\circ}$ in longitude (corresponding to a 3D separation of 0.0416~au or ${\sim}$8.95\,$R_{\odot}$). We have focussed on characterising and comparing properties of the interplanetary shock, sheath region, and magnetic ejecta as measured by the two spacecraft. Overall, we have found some similarities, but also some profound differences between the two sets of in-situ data that, although resulting from flank encounters with the structure as a whole, are characterised by a relative non-radial angular separation (${\sim}5^{\circ}$) that is smaller than the typical errors associated with 3D reconstructions of CMEs based on remote-sensing data---for example, \citet{verbeke2023} found minimum uncertainties of $6^{\circ}$ in latitude and $11^{\circ}$ in longitude using the GCS model\edit1{, and \citet{kay2023} estimated a typical difference of $4^{\circ}$ in latitude and $8^{\circ}$ in longitude between two independent reconstructions of the same event}.

Ultimately, the findings of this work resonate with the conclusions and recommendations outlined by \citet{lugaz2018}, i.e.\ that the mesoscale range of the parameter space is still left largely unexplored in the context of CMEs, resulting in a knowledge gap over radial separations of 0.005--0.050~au and longitudinal separations of $1^{\circ}$--$12^{\circ}$ (i.e., the expected size of the cross-section of a CME ejecta at 1~au). Even more so, the work reported here shows that the results of \citet{lugaz2018} and \citet{alalahti2020} apply even at shorter heliocentric distances than 1~au, despite the generally lower amount of evolutionary processes such as interactions and deformations that are expected to have taken place. Fortunately, the importance of investigating mesoscales in the solar wind has been gaining more traction in recent years \citep[e.g.,][]{viall2021}, and novel missions are being proposed to investigate variations and variability of solar transient events \citep[e.g.,][]{allen2022, lugaz2023, nykyri2023}. The potential benefits of a constellation of spacecraft designed to explore the mesoscale region would be invaluable not only for improving upon our knowledge of the fundamental physics of CMEs, but also of other transient phenomena such as stream interaction regions and solar energetic particles.

The findings of this work have also implications in space weather research and operations. First of all, we remark on the importance of improving upon the available remote-sensing observations for trajectory and size determination of CMEs, since in-situ measurements can show some prominent variability over short spatial scales. This could be achieved by increasing the number of operational telescopes in space, and/or by observing the Sun and its environment from novel viewpoints such as from outside the ecliptic plane \citep[e.g.,][]{gibson2018, deforest2023, howard2023, palmerio2023a}. From a modelling and forecasting perspective, we highlight the importance of considering ``swarms'' of virtual spacecraft (around Earth or the target body of interest) when generating predictions, with separations of the synthetic probes even as little as $5^{\circ}$ in non-radial distance to properly account for uncertainties in the physical processes that are being simulated as well as in the determination of CME input parameters from observations \citep[e.g.,][]{scolini2019, asvestari2021, maharana2023, palmerio2023b}. Additionally, this work showcased the need for reaching a better understanding and improving predictions of CME-driven sheath regions---an effort that is only in its infancy at the time of writing \citep[e.g.,][]{kay2020}.

The relevance of this work to space weather research is not limited to improving predictions at Earth, but has also applications for planetary science. The event studied here was encountered in situ by two spacecraft at Mercury's orbit, allowing us to analyse its characteristics at heliocentric distances that are significantly closer to the Sun than 1~au. Understanding the structure and evolution of CMEs within Mercury's orbit, in fact, is of high importance for the success of Bepi after its planned Mercury orbital insertion, as solar eruptive transients are the sources of many of the planetary dynamics on which the mission will focus \citep[e.g.,][]{milillo2020}. Our knowledge of space weather at Mercury, including solar wind interaction with the Hermean magnetosphere, is at the moment rather limited, although some progress has been achieved via data collected during a few flybys (e.g., from Mariner~10) and orbital measurements from MESSENGER \citep[e.g.,][]{killen2004, winslow2017}. Future observations from Parker, SolO, and Bepi (cruise phase) of CMEs at Mercury's orbit in the solar wind, as well as from Bepi at Mercury after orbital insertion, will greatly enhance our understanding of space weather phenomena at the first planet from the Sun.

Finally, this analysis has been possible due to a fortuitous relative configuration of two probes, Bepi and Parker, at a time during which a remarkable CME was fortuitously launched in their direction. Even though such events are understandably rare, taking advantage of these ``special kinds'' of multi-spacecraft encounters, i.e.\ characterised by small radial \emph{and} angular separations, is a concrete strategy that can be applied at present, at least whilst we wait for one or more dedicated missions to help characterising the solar wind and its transients over all the relevant spatial scales.


\section*{Acknowledgments}

E.P.\ acknowledges support from NASA's Parker Solar Probe Guest Investigators (PSP-GI; no.\ 80NSSC22K0349), Heliophysics Theory, Modeling, and Simulations (HTMS; no.\ 80NSSC20K1274), and Living With a Star Strategic Capabilities (LWS-SC; no.\ 80NSSC22K0893) programmes, as well as the Parker Solar Probe WISPR contract no.\ NNG11EK11I to NRL (under subcontract no.\ N00173-19-C-2003 to PSI).
F.C., S.P., and A.J.W.\ acknowledge the financial support by an appointment to the NASA Postdoctoral Program at the NASA Goddard Space Flight Center, administered by USRA through a contract with NASA.
B.S.-C.\ acknowledges support from the UK-STFC Ernest Rutherford fellowship ST/V004115/1 and the BepiColombo guest investigator grant ST/Y000439/1.
B.J.L.\ acknowledges support from NASA HSR no.\ 80NSSC20K1448 and HGI no.\ 80NSSC21K0731. B.J.L.\ and C.O.L.\ acknowledge NASA LWS grant no.\ 80NSSC21K1325.
D.L.\ acknowledges support from NASA Living With a Star (LWS) programme NNH19ZDA001N-LWS.
A.N.Z., L.R., and M.M.\ thank the Belgian Federal Science Policy Office (BELSPO) for the provision of financial support in the framework of the PRODEX Programme of the European Space Agency (ESA) under contract numbers 4000112292, 4000134088, 4000134474, and 4000136424.
L.R.-G.\ acknowledges the financial support by the Spanish Ministerio de Ciencia, Innovaci{\'o}n y Universidades FEDER/MCIU/AEI Projects ESP2017-88436-R and PID2019-104863RB-I00/AEI/10.13039/501100011033. L.R.-G.\ and N.D.\ acknowledges support by the European Union's Horizon 2020 research and innovation program under grant agreement no.\ 101004159 (SERPENTINE).
N.D.\ also acknowledges funding of the Academy of Finland (SHOCKSEE, grant no.\ 346902).

BepiColombo is an ESA--JAXA mission, where MPO has been built and is operated by ESA.

Parker Solar Probe was designed, built, and is now operated by the Johns Hopkins Applied Physics Laboratory as part of NASA's LWS programme (contract no.\ NNN06AA01C). Support from the LWS management and technical team has played a critical role in the success of the Parker Solar Probe mission. 

Solar Orbiter is a space mission of international collaboration between ESA and NASA, operated by ESA. The EUI instrument was built by CSL, IAS, MPS, MSSL/UCL, PMOD/WRC, ROB, LCF/IO with funding from the Belgian Federal Science Policy Office; the Centre National d’Etudes Spatiales (CNES); the UK Space Agency (UKSA); the Bundesministerium f{\"u}r Wirtschaft und Energie (BMWi) through the Deutsches Zentrum f{\"u}r Luft- und Raumfahrt (DLR); and the Swiss Space Office (SSO).

\facilities{Bepi (MPO-MAG); GOES (SUVI); Parker (FIELDS, SWEAP, WISPR); SOHO (LASCO); SolO (EUI); STEREO (SECCHI)}
\software{AstroPy \citep{astropy2022}; SolarSoft \citep{freeland1998}; SunPy \citep{sunpy2020}}


\appendix

\section{CME Observations by the WISPR Cameras Onboard Parker} \label{app:wispr}

Observations of the 2022 February 15 CME captured by the WISPR Inner (field of view from $13.5^{\circ}$ to $53.0^{\circ}$) and Outer (field of view from $50.0^{\circ}$ to $108.0^{\circ}$) cameras onboard Parker are part of the observing sequence scheduled for the spacecraft's Encounter~11 (E11) with the Sun, with perihelion on 2022 February 25 at 13.3\,$R_{\odot}$. An overview of the Sun--WISPR observing geometry, available data, and science achievements during E11 has been provided by \citet{liewer2023}.

\begin{figure*}[th!]
\centering
\includegraphics[width=\linewidth]{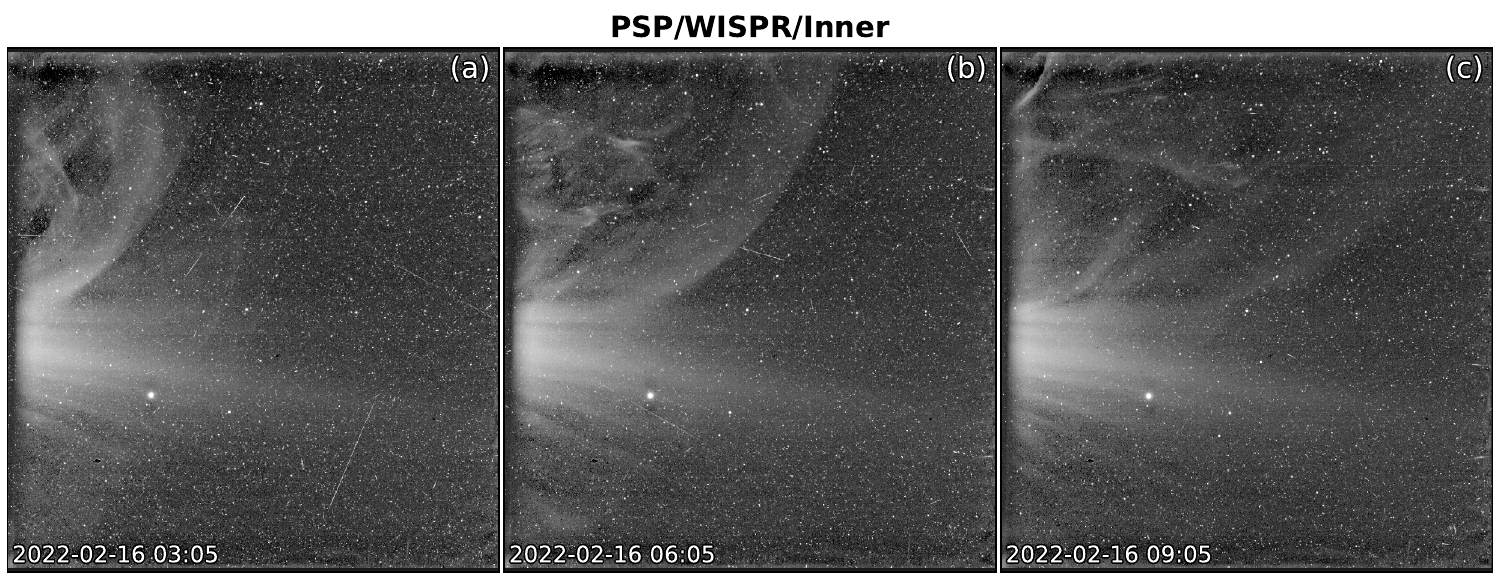}
\caption{Overview of the observations of the 2022 February 15 event taken by the WISPR Inner camera onboard Parker. The CME is shown as it propagates in three frames separated by 3~hr each. The planet (brightest point) in the lower-left quadrant of each image is Earth.}
\label{fig:wispr}
\end{figure*}

A few images from the WISPR/Inner camera containing the 2022 February 15 CME in its field of view are shown in Figure~\ref{fig:wispr}. These data were processed via the L3 Algorithm developed by the WISPR instrument team and described in \citet{liewer2023}. During our period of interest (2022 February 15--17, i.e.\ approximately 10~days before the E11 perihelion), WISPR was pointing over the western limb of the Sun (from Parker's viewpoint), as can be seen by the streamer emanating from the left of each frame in Figure~\ref{fig:wispr}. The CME was observed to cross the telescope's field of view from left to right predominantly over its top half, indicating that Parker observed a CME that propagated mostly towards the north and experienced little-to-no deflection towards the equatorial plane after leaving the Sun, confirming the assessment in Section~\ref{sec:erupt} based on other available remote-sensing imagery.

As mentioned in Section~\ref{subsec:shock}, the CME-driven shock impacted Parker on February 16 at 07:25~UT, i.e.\ between the observation times of panels (b) and (c) of Figure~\ref{fig:wispr}. This means that the spacecraft imaged the CME as it was being impacted by it, an occurrence historically considered exceptional and that is becoming more common during the Parker era \citep[e.g.,][]{romeo2023, wood2023} as the probe approaches the Sun by lowering its perihelion and the Solar Cycle 25 activity levels increase. Shortly after the time of the ejecta leading edge passage, i.e.\ February 16 at 15:18~UT, trailing flows from the CME were observed in the WISPR/Outer camera---these images, however, are characterised by high noise levels and moving features in them can be better appreciated in video format\footnote{WISPR movies for the Inner and Outer cameras, sorted chronologically by solar encounter number, are available at \href{https://wispr.nrl.navy.mil/wisprdata}{https://wispr.nrl.navy.mil/wisprdata}.}, hence they are not shown here.

\section{Processing and Calibration of the Bepi Magnetic Field Data} \label{app:mpomag}

The primary objective of the magnetometer instrument MPO-MAG onboard the Mercury Planetary Orbiter (MPO) of the BepiColombo mission is to map Mercury's magnetic field as well as the interaction of the hermean magnetosphere with the solar wind. In order to characterise the solar wind at short distances from the Sun (i.e., at Mercury’s orbit), MPO-MAG has been in operation during most of the Bepi cruise phase to Mercury, contributing to studies of solar wind turbulence and transient phenomena \citep{heyner2021}.
 
Since launch, MPO-MAG has been in continuous operation outside the solar electric propulsion periods \citep[SEPS;][]{montagnon2021}, until beginning of 2022 when it was decided that the instrument will remain in operation during these periods in background mode. MPO-MAG observations during arcs remain of high accuracy but special pre-processing cleaning by the instrument team is needed as the two sensors of the instrument are influenced by the magnetic disturbance field of the spacecraft. The 2022 February 15 event occurred during one of these SEPS arcs, starting on 2022 February 15 at 02:17:20~UT. The main consequences dependent on the individual thrusters being used and the thrust levels, and a shift of the components by 10~nT was found, as well as an increase in the noise floor. In this regard, the MPO-MAG team cleaned the data phases and removed the noise for this study, by \edit1{low-pass filtering} the field as well as \edit1{linearly interpolating the time-dependent instrument offsets before and after the event studied in this work. The quality criterion adopted here is the sensor component differences, which should be zero in case of well-cleaned data.}
 
We consider that the version of the data used in this study is the most accurate available for this event and the total field and components can be trusted within an error of 10~nT in magnitude. We also note that the arrival times of the shock and sheath match very well with the Bepi high-energy particles observed by the BepiColombo Environment Radiation Monitor \citep[BERM;][]{pinto2022}, the only other instrument in operation during this SEPS arc \edit1{\citep[these measurements, not shown here, are presented in detail by][]{khoo2024}}. Therefore, we also trust that the timing of the crossing of the different regions of the CME observed by MPO-MAG are accurate.

\section{Flux Rope Fits for the Magnetic Obstacle Interval at Parker} \label{app:psp_fits}

\begin{figure}[th!]
\centering
\includegraphics[width=.5\linewidth]{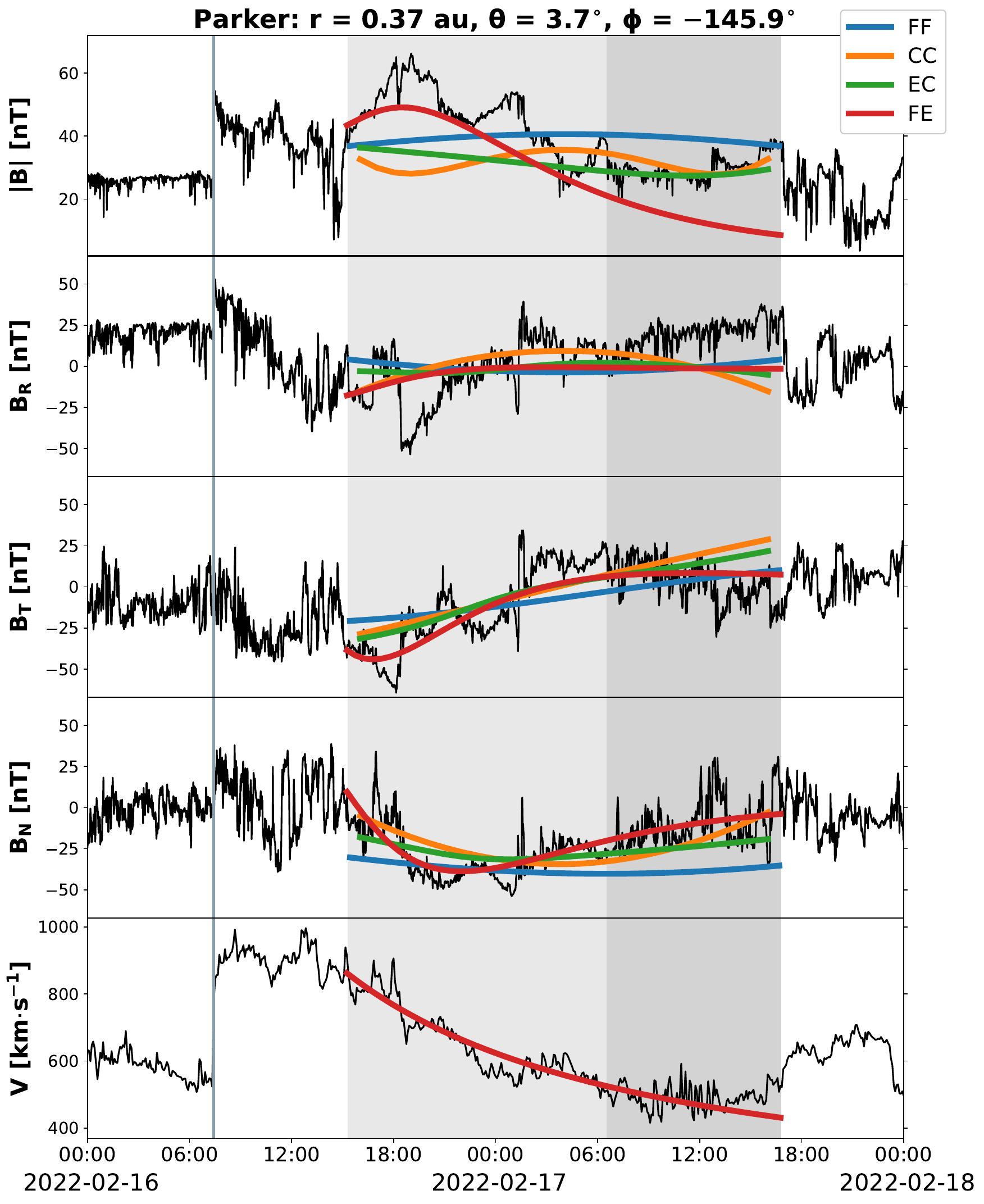}
\caption{Flux rope fitting results at Parker obtained when considering the full magnetic ejecta interval, showing magnetic field and bulk speed data as well as the corresponding quantities resulting from different flux rope models.}
\label{fig:psp_fits}
\end{figure}

To compare the flux rope fitting results at Bepi with both ejecta interval possibilities at Parker, we repeat the reconstructions shown in Figure~\ref{fig:fr_fits}(b) for the magnetic cloud interval by considering the full magnetic obstacle identified at Parker. These results are presented in Figure~\ref{fig:psp_fits} and Table~\ref{tab:psp_fits}. \edit1{As in Figure~\ref{fig:fr_fits}(b), we adopt here all the four models considered for the magnetic cloud interval at Parker, i.e.\ FF, CC, EC, and FE. For the first three fitting procedures we employ an average solar wind speed of 589~km$\cdot$s$^{-1}$, whilst for the FE technique we use the fitted speed profile shown in the bottom panel of Figure~\ref{fig:psp_fits},  obtained via a second-order polynomial fit to the corresponding Parker data.}

\begin{table}[th!]
\caption{Flux rope fitting results for the full ejecta interval at Parker. \label{tab:psp_fits} \vspace*{.1in}}
\centering
\renewcommand{\arraystretch}{1.05}
\begin{tabular}{l@{\hskip .5in}cccc}
\toprule
& \textbf{FF} & \textbf{CC} & \textbf{EC} & \textbf{FE}\\
\midrule
$H$ & +1 & +1 & +1 & +1 \\
$\vartheta_{0}$ [$^{\circ}$] & $-7.1$ & $-49.8$ & $-51.4$ & $-59.7$ \\
$\varphi_{0}$ [$^{\circ}$] & 181.2 & 1.0 & 326.5 & 312.6 \\
$B_{0}$ [nT] & 71.0 & 41.6 & 28.2 & 83.7 \\
$R_{0}$ [au] & 0.053 & 0.135 & 0.224 & 0.120 \\
$p_{0}$ [$R_{0}^{-1}$] & 0.905 & 0.485 & 0.307 & 0.430 \\
$\tau$ [au$^{-1}$] & 16.32 & 0.84 & 1.79 & 7.21 \\
$\Phi_\mathrm{ax}$ [10$^{21}$ Mx] & 0.60 & 2.92 & 2.21 & 3.68 \\
$\Phi_\mathrm{po}$ [10$^{21}$ Mx/au] & 3.50 & 4.43 & 5.72 & 9.37 \\
C1 & --- & 1.30 & 0.65 & ---\\
$\psi$ [$^{\circ}$] & --- & --- & 122.8 & --- \\
$\delta$ & --- & --- & 0.41 & --- \\
\bottomrule
\end{tabular}
\vspace*{.1in}
\begin{tablenotes}
\item \emph{Notes.} The parameters shown are in the same format as in Table~\ref{tab:fr_fits}. The assumed average speed for the CME ejecta in the FF, CC, and EC fits is 589~km$\cdot$s$^{-1}$, whilst the FE procedure uses the time-dependent speed shown in the bottom panel of Figure~\ref{fig:psp_fits}.
\end{tablenotes}
\end{table}

In the case of the magnetic obstacle interval at Parker, the different flux rope fitting techniques appear visually significantly less consistent with each other than the ones shown in Figure~\ref{fig:fr_fits} for both the ejecta at Bepi and the magnetic cloud at Parker. This may be due to the strong asymmetry seen in the magnetic field magnitude between the leading and trailing portions of the magnetic obstacle. The inconsistencies are also reflected in the resulting parameters presented in Table~\ref{tab:psp_fits}, which include e.g.\ flux rope axis directions that differ by up to ${\sim}53^{\circ}$ in latitude and ${\sim}180^{\circ}$ in longitude, as well as impact parameters that range between ${\sim}30$\% and ${\sim}90$\% away from the central axis along the cross-sectional radius.

Visually, the FE fit appears to better match the data, which is not surprising since this model takes CME expansion into account and the 2022 February 15 event was strongly expanding when it was measured in situ by Parker (as shown in the bottom panel of Figure~\ref{fig:psp_fits}). However, even the FE results do not reproduce the $B_{R}$ profile particularly well, which on the other hand is poorly fitted by all models even when considering the magnetic cloud interval only (see Figure~\ref{fig:fr_fits}(b)). Overall, each technique applied to the full magnetic obstacle at Parker retrieves a flux rope structure that is fundamentally different from the corresponding results obtained from the magnetic cloud time series, showcasing the complexity of the structure under analysis.

\bibliographystyle{aasjournal}
\bibliography{bibliography}

\end{document}